\documentclass[pra,floatfix,superscriptaddress,showpacs,twocolumn,nofootinbib]{revtex4}

\usepackage{amsfonts}
\usepackage{amssymb}
\usepackage{amsmath}
\usepackage{graphics}
\usepackage[pdftex]{graphicx}

\newcommand{\ket}[1]{{|{#1}\rangle}}

\begin{document}


\title{Classical Processing Requirements for a Topological Quantum Computing System.}
\author{Simon J. Devitt}
\email[email:]{devitt@nii.ac.jp}
\affiliation{National Institute of Informatics, 2-1-2 Hitotsubashi, Chiyoda-ku, Tokyo-to 101-8430, Japan}
\author{Austin G. Fowler}
\affiliation{Center for Quantum Computing Technology, University of Melbourne, Victoria 3010, Australia}
\author{Todd Tilma}
\affiliation{National Institute of Informatics, 2-1-2 Hitotsubashi, Chiyoda-ku, Tokyo-to 101-8430, Japan}
\author{W. J. Munro}
\affiliation{Hewlett Packard Laboratories, Bristol BS34 8HZ, United Kingdom}
\affiliation{National Institute of Informatics, 2-1-2 Hitotsubashi, Chiyoda-ku, Tokyo-to 101-8430, Japan}
\author{Kae Nemoto}
\affiliation{National Institute of Informatics, 2-1-2 Hitotsubashi, Chiyoda-ku, Tokyo-to 101-8430, Japan}

\date{\today}

\begin{abstract}
Dedicated research into the design and construction of a large scale Quantum Information 
Processing (QIP) system is a complicated task.  The design of an experimentally 
feasible quantum processor must draw upon results in multiple fields; from experimental 
efforts in system control and fabrication through to far more abstract areas such as quantum 
algorithms and error correction.   Recently, the adaptation of topological coding models 
to physical systems in optics has illustrated a possible long term pathway to truly large scale QIP.  
As the topological model has well defined protocols for Quantum Error Correction (QEC) built in 
as part of its construction, a more grounded analysis of the {\em classical} processing requirements 
is possible.  In this paper we analyze the requirements for a classical processing 
system, designed specifically for the topological cluster state model.  We demonstrate that via 
extensive parallelization, the construction of a classical ``front-end" system capable of processing 
error correction data for a large topological computer is possible today.
\end{abstract}


\maketitle

\section{Introduction\protect}\label{Introduction}

The design and construction of a feasible large scale quantum computing system has 
been a highly sought after and long term goal of quantum information science ever since 
the first physical system proposals were made in the mid 1990's~\cite{CZ95,CFH97,GC97,K98,LD98,KLM01,MOLTW99,NPT99,ARPA,KMNRDM07}.  While experimental 
advances in quantum computing have been pronounced~\cite{CNHM03,YPANT03,CLW04,HHB05,GHW05,G06,H06,G07,OPWRB03} we are not yet at the 
stage where we can faithfully claim a multi-million qubit device is just around the corner. 

Nevertheless, in order for experimental progress to be made, the fundamental theoretical 
building blocks for a large scale computer need to be firmly in place.  This theoretical 
development is not restricted to the discovery of new protocols for computation, algorithms or error 
correction but it also includes the architectural engineering of future computers.  

While there has been steady progress over the past 15 years on designing novel and (more 
importantly) experimentally feasible large scale processor architectures, 
the complication of implementing appropriate and efficient error correction procedures and 
designing systems which can trivially be scaled to the level of a ``programmable", multi-task 
computer is still a daunting and often neglected area of research.  

Recently, the introduction of theoretical ideas such as topological cluster state quantum  
computing (TCQC)~\cite{RHG07,RH07,FG08} and the single photon optical 
architecture~\cite{DFSG08,IM09} 
gives us an idea of what a truly large scale device may possibly look like.  
The modular design of the cluster preparation network, and the measurement based nature of 
the computational model, gives this design something that other architectures arguably lack, 
a strictly modular scaling of the entire computer.  

While the design introduced in Refs.~\cite{DFSG08,DMN08} 
is not necessarily the optimal way to construct a 
large scale quantum computer, it does contain several key elements, 
easing the conceptual design of a large scale computer.  For example:
\begin{enumerate}
\item Utilizing a computational model that is fundamentally constructed from error 
correction, rather than the implementation of codes on top of an otherwise independent computational 
model.
\item Having a modular construction to the computer.  The fundamental quantum component 
(the photonic chip~\cite{SEDG08,DFSG08}), is a comparatively simple quantum device.  Scaling the computer 
arbitrarily requires the addition of more chips in a regular and known manner.  
\item Employing a computational model exhibiting high fault-tolerance thresholds~\cite{RHG07}, 
which relieves the pressure on experimental fabrication and control.
\item Utilizing a measurement based model for computation~\cite{RB01}. By employing 
a measurement based computational model, the quantum component of the 
computer is a simple state preparation network.  Therefore, programming such a device 
is a problem of classical software, not of hardware.
\end{enumerate}
These properties, as well as others, allowed us to consider the structure of an extremely large 
mainframe-type device.  In Ref.~\cite{DMN08}, the quantum analogue of high performance computing 
was examined.  The conceptual scalability of the optical architecture allowed us to 
examine the operating conditions, physical structure and resource costs of a computer employing 
extensive topological error correction to the level of 2.5 million {\em logical} qubits.  

In addition to examining the implementation of a large scale TCQC, 
the nature of the topological model also allows for a more concrete discussion on an 
often neglected, but important aspect of quantum information processing, namely what are the 
classical computational requirements of a large scale device?  In this paper we attempt to answer 
this question.  

There have been several broad investigations into the classical structure, 
design and operation of a large scale quantum computer~\cite{VM06,S07}, 
but investigation into this topic is difficult.  The primary 
obstacle in analyzing classical requirements is that the quantum architecture generally 
has to be specified.  
Since all classical processing is ultimately dependent on both the computational model 
employed at the quantum level and more importantly the error correction protocols utilized, 
a detailed analysis of the classical front end must wait for the design of the quantum processor.  

In this paper we specifically analyze the classical front end requirements to perform active error 
correction on a 3D topological cluster lattice prepared by the photonic chip network.  
This analysis will be restricted to the classical 
system required to implement the underlying error correction procedures in the topological 
model, without  the execution of an active quantum algorithm.  

Although we present this analysis in the context of the optical network presented in Ref.~\cite{DFSG08}, 
it should be stressed that this analysis is still highly relevant for any physical architecture employing 
the 2D or 3D topological model~\cite{BK01, DKLP02,RH07,RHG07,FSG08}.  
Our analysis demonstrates that with several 
optimizations of the classical processing and the ability to significantly parallelize
classical error correction processing, the classical computational requirements for large scale 
TCQC are indeed within the capabilities of {\em today's} processing technology.   

Section~\ref{sec:preliminaries} very briefly reviews the nature of the topological cluster 
model in order to fully specify what is required of the classical processing.  
Section~\ref{sec:flow} reviews the flowing nature of the preparation network, and 
how this relates to the optical measurement layer, and the first level of classical 
processing.  In Section~\ref{sec:classical} we overview the basic requirements of the classical 
network and how target processing rates are related to the clock cycle of the quantum 
network.
In Section~\ref{sec:benchmarking} we introduce classical benchmarking data for 
the minimum weight matching algorithm utilized for error correction and illustrate, given this data, 
how error correction processing can be parallelized over the entire computer.  We conclude by 
illustrating how the parallelization of the classical processing allows, in principal, for the 
construction of a classical error correcting front end for large scale TCQC
with classical processing technology available today.

\section{Topological Error Correction in the Optical Architecture}
\label{sec:preliminaries}

TCQC was first introduced by Raussendorf, Harrington and Goyal in 
2007~\cite{RH07,RHG07}.  This model incorporates the ideas stemming from 
topological quantum computing 
introduced by Kitaev~\cite{K97} and cluster state computation~\cite{RB01}, leading to a very attractive 
computational model incorporating error correction by construction and exhibiting a high 
threshold error rate.  

As with any measurement based computational model, computation 
proceeds via the initial construction of a highly entangled multi-qubit state.   Fig.~\ref{figure:cell} illustrates 
the structure of the cluster.  Each node in the cluster represents a physical qubit, initially prepared in 
the $\ket{+} = (\ket{0}+\ket{1})/\sqrt{2}$ state, and each edge represents a controlled-$\sigma_z$ 
entangling gate between qubits.  This is the fundamental unit cell of the cluster, which 
repeats in all three dimensions.  

Computation under this model is achieved via the consumption of the cluster along one of the 
three spatial dimensions~\cite{FG08} (simulated time).  {\em Logical}
qubits are defined via the creation of ``holes" or ``defects" 
within the global lattice and multi-qubit operations are achieved via braiding 
(movement of these defects around one another) as the cluster is consumed along the direction 
of simulated time.  The specific details for computation under this model are not important for this 
discussion and we encourage the reader to refer to Refs.~\cite{RHG07,FG08} for further details.  
For this analysis, the effect of errors on a topological lattice is the important factor.

\subsection{Error effects}
\begin{figure}[ht]
\begin{center}
\resizebox{80mm}{!}{\includegraphics{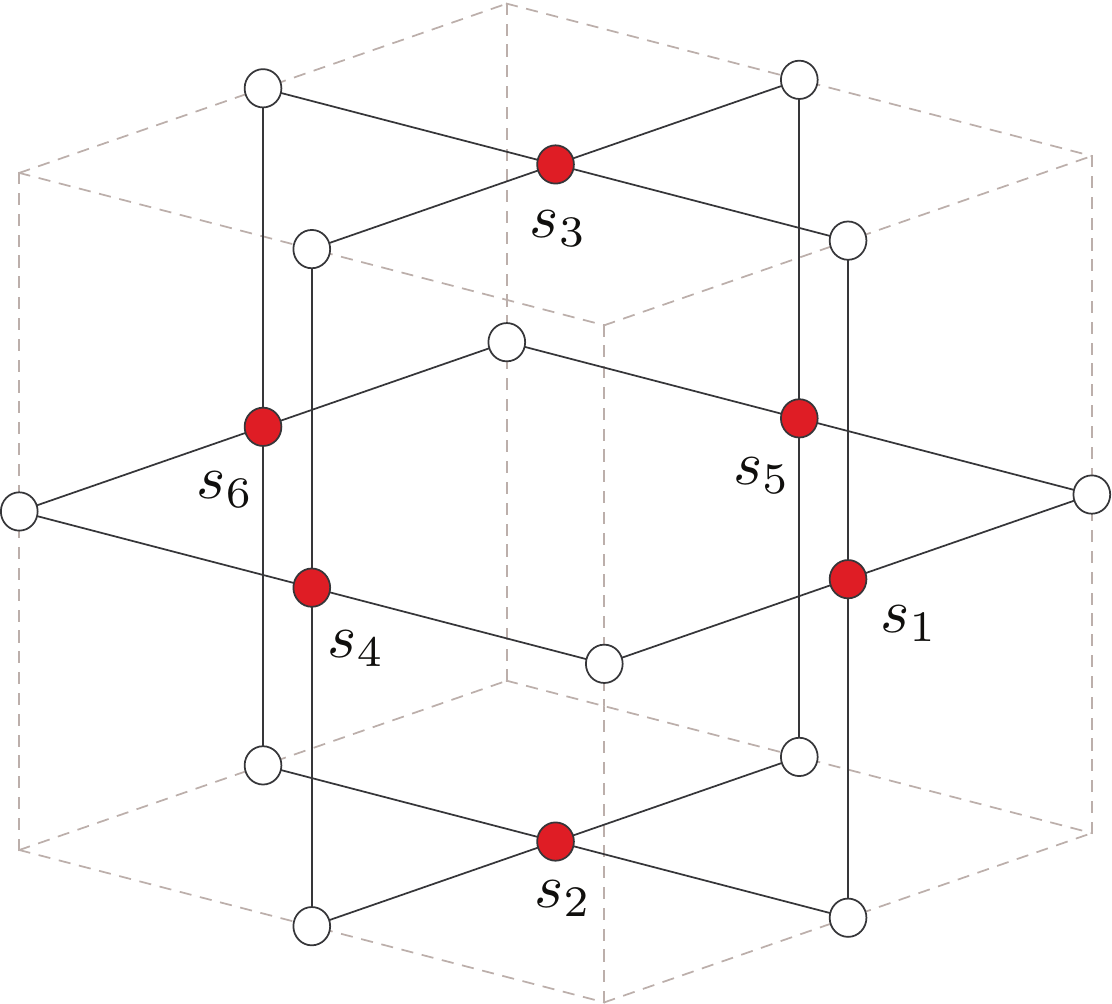}}
\end{center}
\vspace*{-10pt}
\caption{Unit cell of the 3D cluster, which extends arbitrarily in all three dimensions.  In the 
absence of errors, the entanglement generated within the lattice sets up correlation conditions 
when certain qubits are measured.  For each unit cell of the cluster, quantum correlations 
guarantee that the measurement results of the six face qubits (shown above) 
return an even parity result [Eq.~\ref{eq:one}].}
\label{figure:cell}
\end{figure}

Quantum errors in this model manifest in a very specific way.  In Fig.~\ref{figure:cell}, illustrating the 
unit cell of the cluster, we have illustrated six face qubits shown in red.  If no errors are 
present in the system, measuring each of these six qubits in the 
$\ket{+}$ or $\ket{-}$ states ($\sigma_x$ basis) will return an even parity result.  If 
we denote the {\em classical} result of these six measurements as $s_i \in \{0,1\}$, $i = 1,..,6$, then
\begin{equation}
(s_1+s_2+s_3+s_4+s_5+s_6) \text{ mod } 2 = 0.
\label{eq:one}
\end{equation}
This result is a consequence of the quantum correlations established in the preparation of the cluster.  
If the cluster is prepared perfectly, these six qubits are placed into a state that is a $+1$ 
eigenstate of the operator
\begin{equation}
K = \sigma_x^1 \otimes \sigma_x^2 \otimes \sigma_x^3 \otimes \sigma_x^4 \otimes \sigma_x^5 \otimes \sigma_x^6,
\end{equation}
where $\sigma_x$ is the Pauli bit-flip operator.  The measurement of each of these six qubits in the 
$\sigma_x$ basis will produce random results for each individual qubit.  However the 
eigenvalue condition of this correlation 
operator guarantees the classical measurement results satisfy Eq.~\ref{eq:one} in the 
absence of errors.

\begin{figure}[ht]
\begin{center}
\resizebox{80mm}{!}{\includegraphics{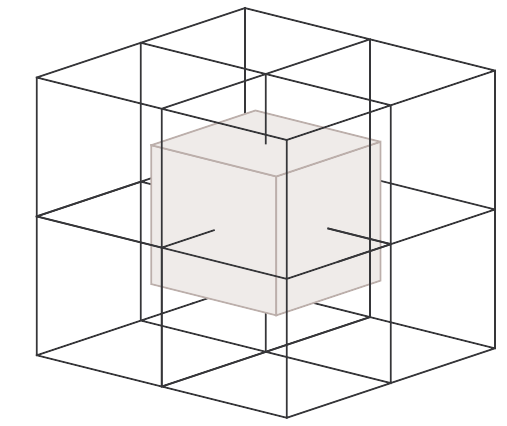}}
\end{center}
\vspace*{-10pt}
\caption{(From Ref.~\cite{FG08}) 
The regular structure of the 3D cluster results in a primal and dual lattice structure.  A 
set of eight unit cells arranged in a cube results in a complete unit cell present at the 
intersection of these eight cells.  Hence there are two self-similar lattices (offset by half a unit cell 
diagonally) known as the primal and dual lattice.  These two structures are extremely important 
for computation under the topological model~\cite{RHG07}.  However, in the context of 
this discussion, the measurement results of the additional nine 
qubts in Fig.~\ref{figure:cell} are associated with the parities of bordering dual cells.}
\label{figure:dual}
\end{figure}

The remaining qubits in each unit cell are also measured in the $\sigma_x$ basis, but 
their results are associated with the parity of cells within the dual lattice 
[Fig.~\ref{figure:dual}].  This property of the cluster is 
not important for this discussion.  What is important is that when no quantum algorithm is 
being implemented, every qubit is measured in the $\sigma_x$ basis and is used to calculate the 
parity of their respective cells. 

Due to the structure of this 3D cluster state, all error channels can effectively be mapped 
into phase errors 
(Pauli-$\sigma_z$ operations applied to phyical qubits in the lattice which takes a state, 
$\alpha\ket{0}+ \beta\ket{1} \rightarrow \alpha \ket{0}-\beta\ket{1}$) or physical qubit loss~\cite{FG08}.  
These two distinct channels are processed slightly differently.

\subsection{Error channel one: Phase errors}
We first consider the effect of phase errors, which 
act to flip the parity of a cell of the cluster.  
As the Pauli operators $\sigma_x$ and $\sigma_z$ anti-commute, $\sigma_z\sigma_x = 
-\sigma_x\sigma_z$, if a phase error occurs to one of these six qubits the correlation condition 
of a cell will flip from 
a +1 eigenstate of the operator $K$ to a -1 eigenstate.  If the correlation condition flips to 
a $-1$ eigenstate of $K$, the classical result of the six individual measurements also flips to 
an {\em odd} parity condition, i.e.
\begin{equation}
(s_1+s_2+s_3+s_4+s_5+s_6) \text{ mod } 2 = 1.
\label{eq:two}
\end{equation}
Errors on qubits on the boundary of each unit cell therefore flip the 
parity of the measurement result from even to odd.  Note that the change of parity of any individual cell 
gives us absolutely no information regarding which one of the six qubits experienced the 
physical error. 

The second important aspect of this structure is that any given qubit lies on the boundary 
of two cells in the cluster.  If a given qubit experiences a phase error it will flip the parity result of the 
two adjacent cells.  This allows us to detect which of the six qubits of a given cell experienced an 
error.  If a single cell flips parity, we then examine the parity result of the six adjacent cells.  Assuming 
that only one error has occured, only one of these six adjacent cells will have also flipped parity 
allowing us to uniquely identify the erred qubit.  
\begin{figure*}[ht]
\begin{center}
\resizebox{110mm}{!}{\includegraphics{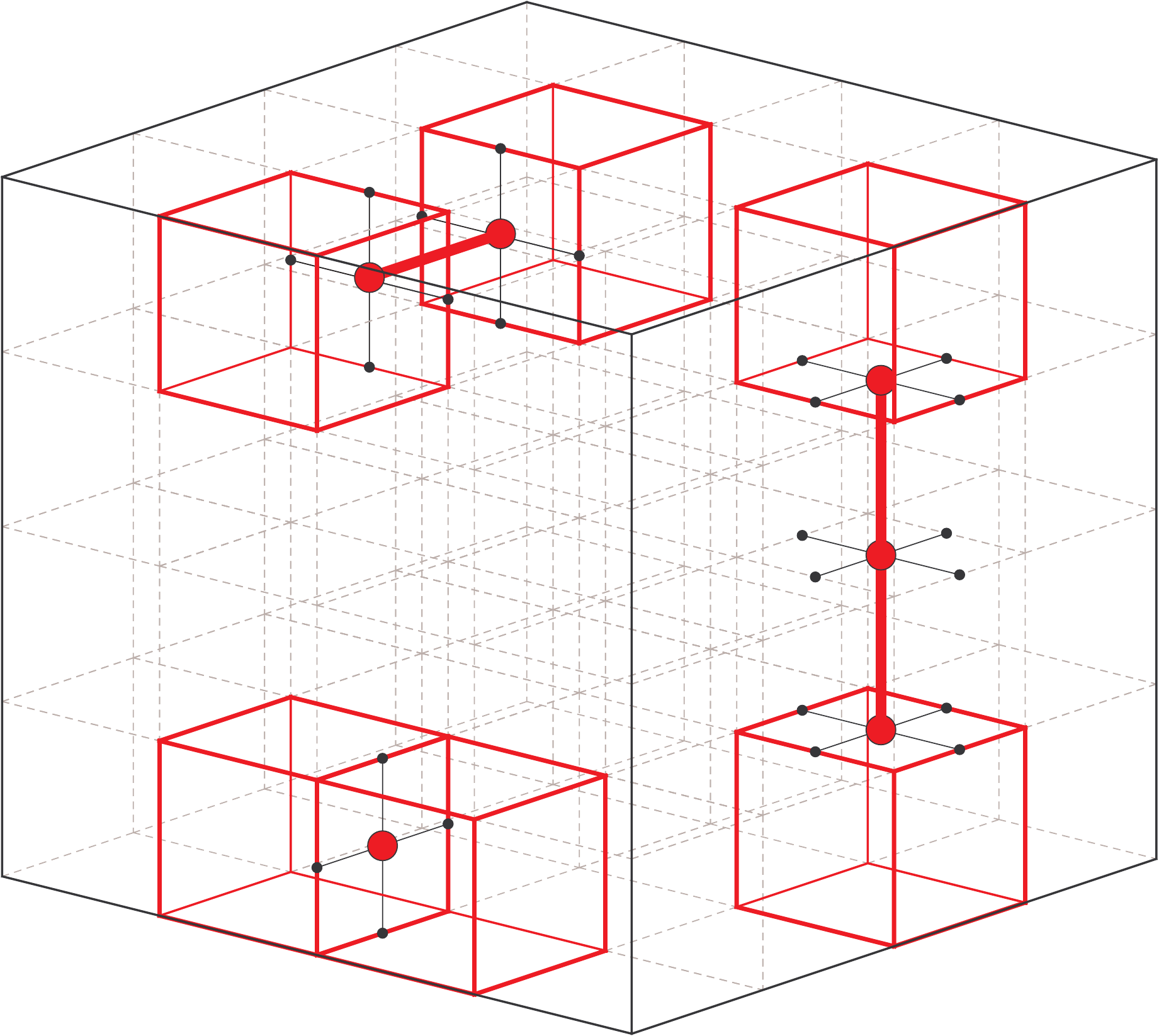}}
\end{center}
\vspace*{-10pt}
\caption{(Taken from Ref.~\cite{FG08})  
Illustration of error effects in the 3D cluster.  Here we show a volume $V=4^3$ of cluster cells 
and the effect of three error chains.  As the parity conditions of Eq.~\ref{eq:one} are 
cyclical (mod 2), the calculation of cell parities only reveals information regarding the endpoints 
of error chains.  Shown above are three examples, one error flips the parity of the two 
cells adjacent to the erred qubit while longer chains only flip the parity of the end point cells.  
The goal of error correction is to faithfully identify these chains given the end point parity data.}
\label{figure:three}
\end{figure*}

If we now consider more than one error within the lattice, we no longer identify the location of individual 
errors but instead identify error chains.  Fig.~\ref{figure:three} from Ref.~\cite{FG08} illustrates.  Here we 
have a 3D cluster consisting of a cube of $4^3$ cells with three error chains.  The first chain 
is a single error which flips the parity of the two adjacent cells, the other two chains illustrate 
the effect of multiple errors.  As the parity conditions are cyclical (mod 2), if two errors occur 
on the boundaries of a given cell the parity result will not change.  Instead, the two cells at the 
endpoints of these error chains are the cells which flip parity.  

Hence, in the TCQC model, it is not the locations of individual errors which are important 
but the endpoints of error chains.  In fact, the symmetries of the cluster do not require 
us to identify the physical error chain corresponding to the detected endpoints.  Once the endpoints 
of the chain are correctly identified {\em all} path of correction operators (Pauli operators which 
are applied to reverse detected errors) which connect the two 
endpoints are equivalent~\cite{RHG07}.  Hence, the goal of error correction in this model is 
to correctly ``pair up" a set of odd parity cells such that the appropriate correction operators can 
be applied.  

Undetectable errors in this model occur when chains become so long that they actually connect 
two boundaries of the lattice.  If a physical error chain completely spans the lattice from one 
boundary to another then each individual cell experiences two physical errors and every cell 
remains in an even parity state.  If the 3D lattice is not used for computation, these error chains 
are actually invariants of the cluster and hence have no effect.  
Once computation 
begins, information is stored by deliberately creating holes (or defects) in this lattice.  These defects 
act as artificial boundaries and consequently error chains connecting defects to {\em any} other boundary 
(either other defects or the boundary of the lattice) are undetectable and cause logical errors 
on stored information.  

From the standpoint of this investigation we are only concerned with performing active error 
correction on a defect free lattice.  We will not be introducing information qubits into the 
cluster.  Instead we will be examining the classical resources required to detect and 
correctly identify error chains in an otherwise perfectly prepared lattice.  

This type of analysis is justified 
as information qubits are essentially regions of the 3D cluster that have simply been removed 
from the global lattice.  Analyzing the classical requirements for the complete, defect free lattice
therefore represents the maximum amount of classical data that needs to be processed for correction. 

\subsection{Error channel two: Qubit loss}
The second major error channel is qubit loss.  As we are motivated by the optical architecture 
introduced in Ref.~\cite{DFSG08}, loss represents a significant error channel.  Unlike other computation 
models for quantum information, the TCQC model can correct loss without additional 
machinery.  Without going into the details, loss events can be modeled by tracing out the qubit that is 
lost from the system and replacing the qubit with the vacuum.  Tracing out the lost qubit 
is equivalent to measuring the qubit in the $\ket{0}$ or $\ket{1}$ state 
(a $\sigma_z$ basis measurement) with the result unknown.  In principal, this type of 
error can be modeled as a standard channel, causing the parity of the respective cells to flip 
with a probability of 50\%.  However, since the qubit is no longer present, we can utilize the vacuum 
measurement to uniquely identify these error events.  

\begin{figure}[ht!]
\begin{center}
\resizebox{80mm}{!}{\includegraphics{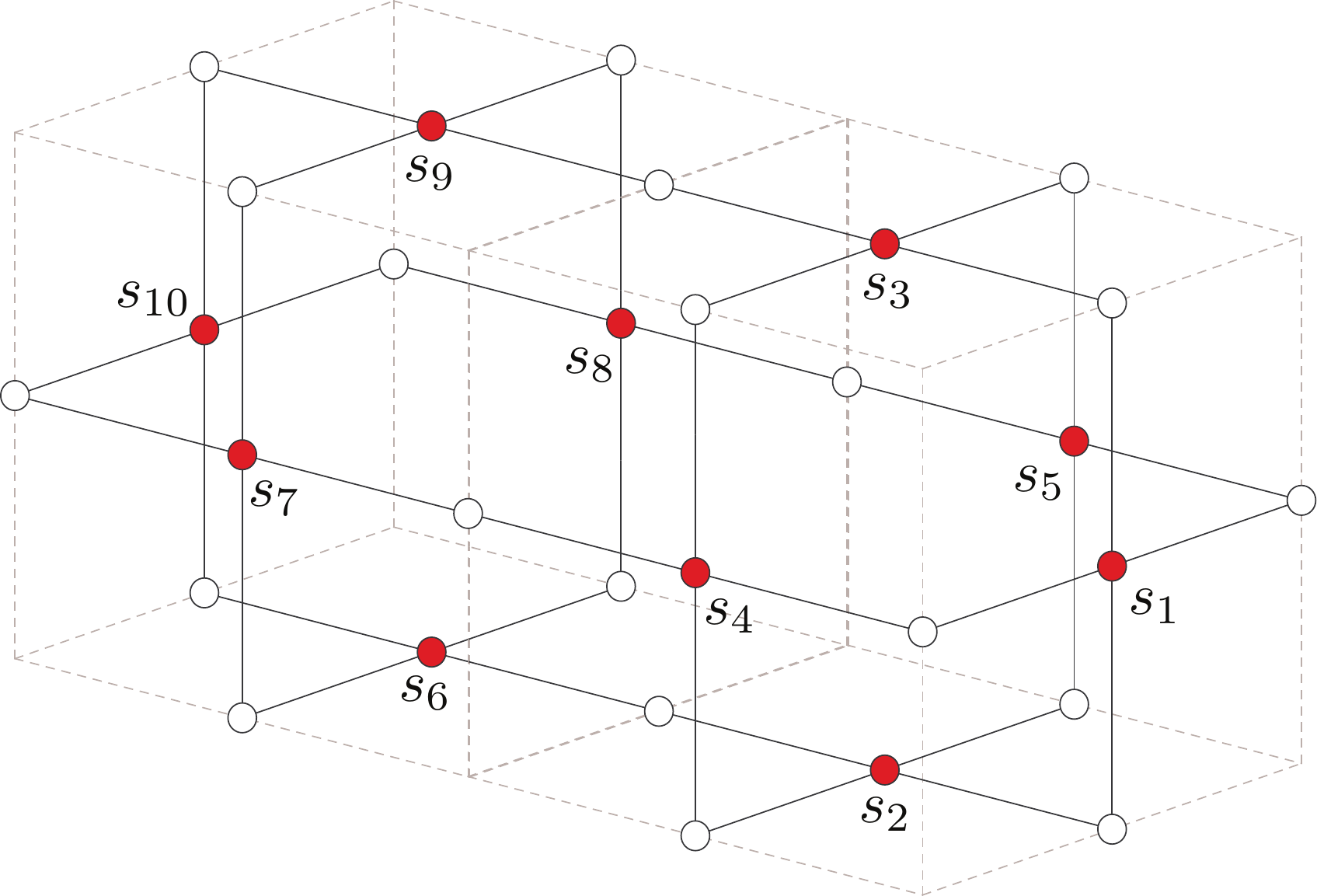}}
\end{center}
\vspace*{-10pt}
\caption{The effect of qubit loss on the parity conditions of the cluster.  When a qubit is lost, it is 
essentially removed from the cluster.  The parity condition of Eq.~\ref{eq:one} is extended to the 
boundary surrounding the loss event.  As a lost qubit is a heralded error (i.e. can be 
detected separately), the parity calculation can be modified to encapsulate 
this larger volume if a loss event is detected.}
\label{figure:loss}
\end{figure}

Illustrated in Fig.~\ref{figure:loss} is the structure of a unit cell when one qubit is essentially measured out 
via loss.  In this case, the boundary of a cell increases around the lost qubit.  Instead of the 
parity conditions being associated with the six face qubits of a given cell, it extends to be 
the combined parity of the ten measurements indicated.  As the loss event is detected via no ``clicks" 
from the detector array, this result is corrected by now taking the parity of this larger structure and 
proceeding as before.  Provided no other errors have occured, the parity of this larger boundary 
will be even, and any additional qubit errors will link this extended cell to a second end point 
with odd parity.  Recent results, obtained in the context of the surface 
code~\cite{DKLP02,FSG08}, have 
demonstrated a high tolerance to heralded loss events~\cite{SBD09}.

\section{The optical computer, a ``flowing" network}
\label{sec:flow}
\begin{figure*}[ht]
\begin{center}
\resizebox{170mm}{!}{\includegraphics{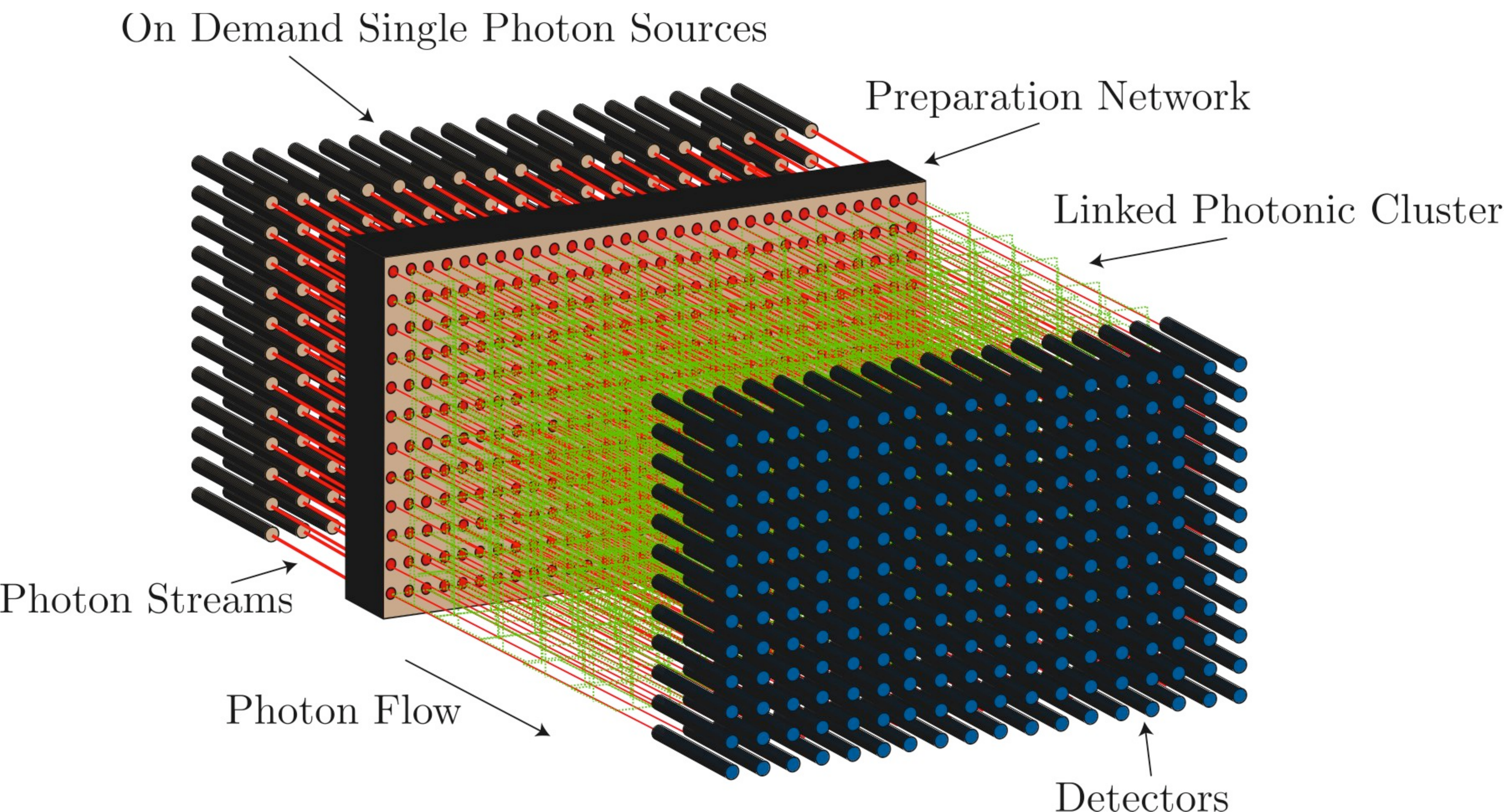}}
\end{center}
\vspace*{-10pt}
\caption{General architectural model for a ``flowing" optical computer.  
Single photon sources inject photonic qubits into 
a preparation network that deterministically links up a 3D 
photonic cluster, layer by layer.  Immediately after the preparation 
network, an array of single photon detectors measures each photon to 
perform computation.  As photons are continually linked to the 
rear of the 3D cluster as the front is consumed, an arbitrarily deep 
3D cluster can be prepared and consumed with finite space.}
\label{figure:flow}
\end{figure*}

As introduced in Ref.~\cite{DFSG08}, optical TCQC can be performed by 
making use of a preparation network of photonic chips~\cite{SEDG08}.  
This network receives a stream of 
single photons (from a variety 
of appropriate sources) and ejects a fully connected topological cluster.  As computation 
proceeds via the successive consumption of lattice faces along the direction of simulated time, 
the preparation network is designed to continuously prepare the lattice along this third dimension 
at a rate equal to the rate of consumption by the detector layer.  Fig.~\ref{figure:flow} illustrates the 
basic design.  

Photons are continuously injected into the rear of the preparation network, ideally from appropriate 
on-demand sources.  Each photon passes through a network of four photonic chips, which act to 
link them together into the appropriate 3D array.  Each photonic chip operates on a fundamental 
clock cycle, $T$, and each chip in the network operates in a well-defined manner, independent 
of the total size of the network~\cite{DFSG08}.  In total, a single photon entering the network at $t=0$ 
will exit at $t=4T$, after which it can be measured by the detector banks.  

Each photonic chip acts 
to entangle a group of five photons into an appropriate state such that the parity condition for 
each cell is satisfied.  After each group of five photons passes through an individual chip, a single atomic 
system contained within each chip is measured and reinitialized, projecting the relevant 
group of 5-photons into an entangled state.  The result of this atomic measurement is fed forward 
to the classical processing layer in order to define a set of initial correlation conditions.  
The cluster is defined such that Eq.\ref{eq:one} is satisfied for all cells.  However, the 
preparation network does not automatically produce these correlation conditions.  
Depending on the measurement 
results of the atomic systems contained within each photonic chip, approximately 50\% of cells 
within the lattice will be prepared with an initial parity condition that is odd.  This can, in principal, be 
corrected to be even for all cells by applying selective single qubit rotations dependent on 
the atomic readout results, but this is unnecessary.  
The initial parity results from the preparation network are simply recorded, endpoints of error chains 
are then identified with cells that have changed parity from this initial state.

As one dimension of the topological lattice is identified as simulated time, the total 2D cross section 
defines the actual size of the quantum computer.  Defects, regions of the cluster 
measured in the $\sigma_z$ basis, are used to define logical qubits and are kept 
well separated to ensure fault tolerance.  The 2D cross section is then continually teleported, via measurement, 
to the next successive layer along the direction of simulated time allowing an algorithm to 
be implemented (in a similar manner to standard cluster state computation~\cite{RB01}).  
\begin{figure*}[ht]
\begin{center}
\resizebox{170mm}{!}{\includegraphics{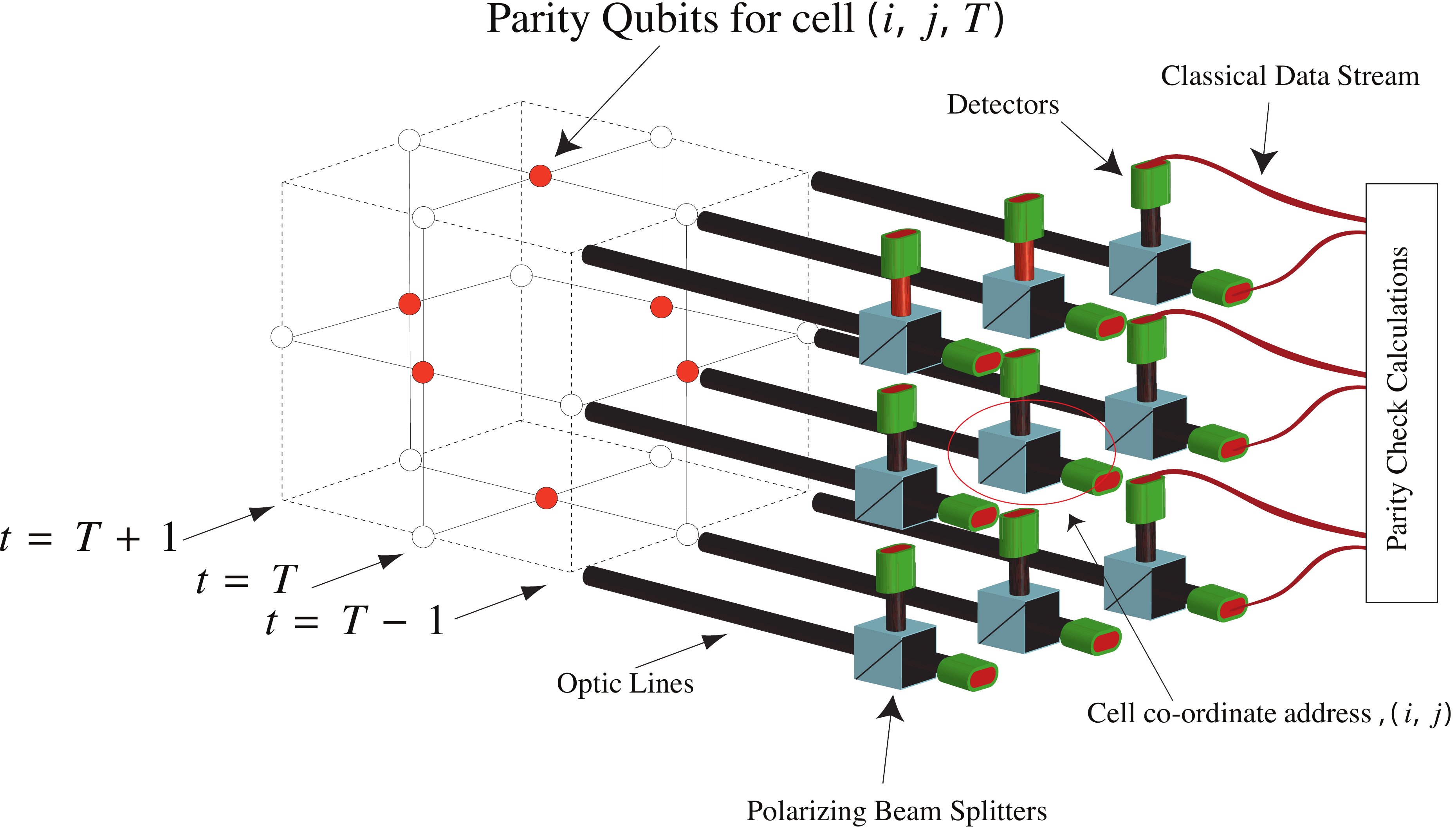}}
\end{center}
\vspace*{-10pt}
\caption{basic structure of the detection layer in the optical network.  A given unit cell of the cluster can be 
associated with 9 optical waveguides containing temporally separated photonic qubits.  For each 
set of 9 detectors (consisting of a polarizing beam splitter and two single photon detectors for 
polarization encoding), 
the central detector is associated with the cross-sectional co-ordinate address, 
$(i,j)$, for the cell.  The temporal co-ordinate, $T$, is associated with the current clock cycle of 
the quantum preparation network.  For each unit cell, in the absence of photon loss, the 
measurement results for the 6 face qubits are sent to the first classical processing layer that 
calculates Eq.~\ref{eq:parity} for each 3D co-ordinate, $(i,j,T)$.  If the parity result differs between 
co-ordinates $(i,j,T-2)$ and $(i,j,T)$, this information is sent to the next classical processing layer.}
\label{figure:detection}
\end{figure*}

In Fig.~\ref{figure:detection} we illustrate the structure of the detection system.  
For the sake of simplicity, 
we are assuming that the basis states for the qubits are photon polarization, $\ket{H} \equiv \ket{0}$ 
and $\ket{V} \equiv \ket{1}$, hence the detection system consists of a polarizing beam splitter (PBS) 
and two single photon detectors.  A given unit cell flows through a set of nine optical lines 
such that the relevant parity is given by, 
\begin{widetext}
\begin{equation}
P(i,j,T) = (s_{(i,j)}^{T-1} + s_{(i-1,j)}^T+s_{(i,j-1)}^T+s_{(i,j+1)}^T+s_{(i+1,j)}^T+s_{(i,j)}^{T+1}) \text { mod } 2 
\label{eq:parity}
\end{equation}
\end{widetext}
where $s_{i,j}^T$ is the detection result $(1,0)$ of detector $(i,j)$ at time $T$.  This result defines 
the parity of the cell $(i,j,T)$ in the lattice.  Loss events would 
result in neither detector firing, at which point the calculation of Eq.~\ref{eq:parity} would be 
redefined based on the loss event to calculate the parity of the boundary around this lost qubit.  

The results of all the detection events are fed directly from the detectors into the first classical 
processing layer.  This layer calculates Eq.~\ref{eq:parity} 
and passes the result forward to the subsequent processing layer if it differs from the initial 
parity.  
This general structure extends across the entire 2D cross section of the lattice with parities 
repeatedly calculated for each unit cell that flows into the detection system.

\section{Classical Processing requirements}
\label{sec:classical}

\begin{figure*}[ht]
\begin{center}
\resizebox{170mm}{!}{\includegraphics{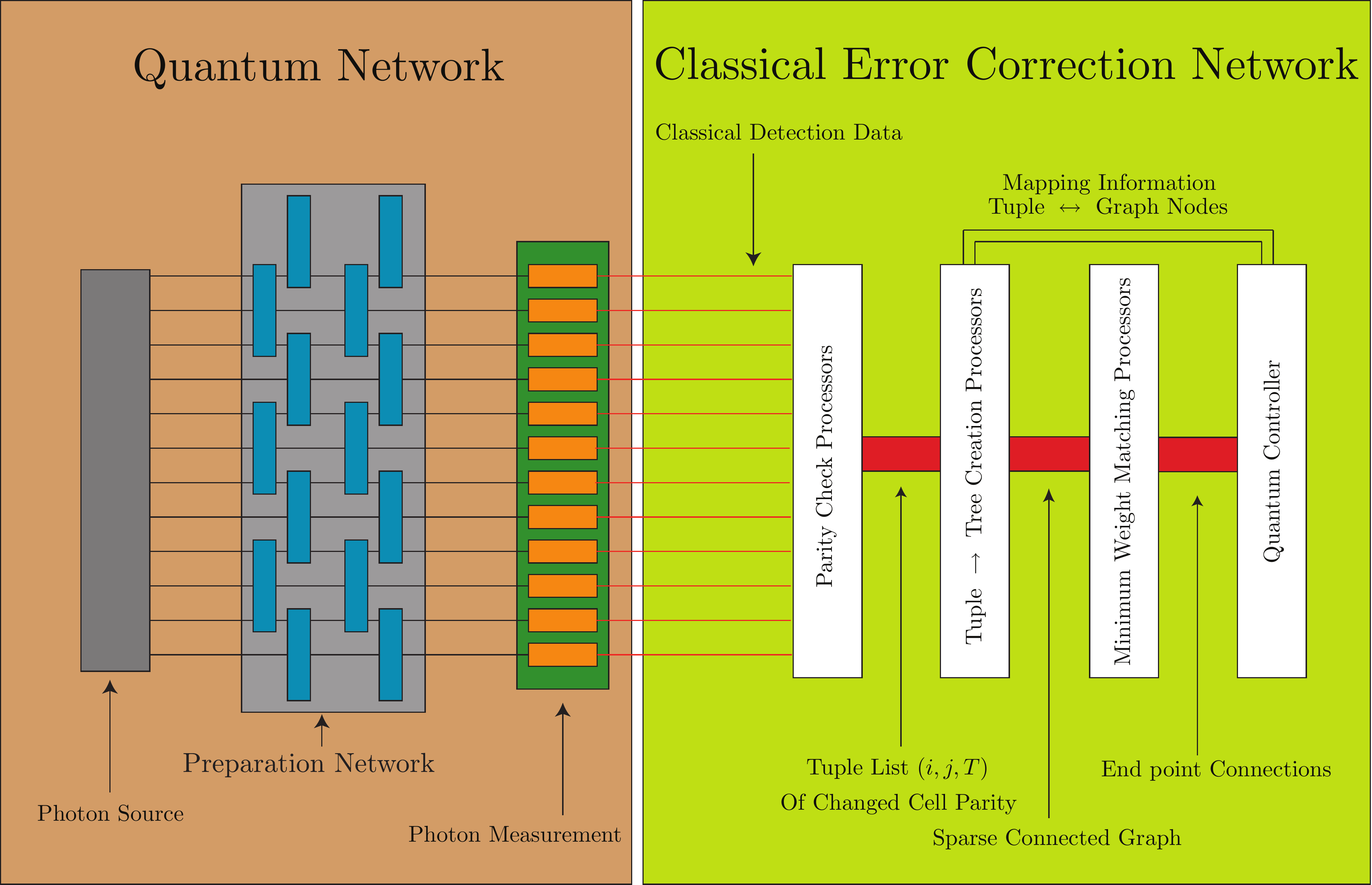}}
\end{center}
\vspace*{-10pt}
\caption{General cross-sectional 
processing structure for the error correction procedures in the optical TCQC.  
The quantum network consists of a bank of single photon sources, photonic chips and 
detectors.  The preparation network continually prepares the cluster along the dimension of 
simulated time.  The detector array will be outputting classical results on the same time 
frame as the quantum clock cycle.  The classical processing layer is required to determine 
likely {\em physical} chains of errors producing the classical parity results detected.  
The classical layer consists of 4 stages, parity check, co-ordinate tuple $\rightarrow$ tree 
creation, minimum weight matching, and the quantum controller.}
\label{figure:structure}
\end{figure*}

Illustrated in Fig.~\ref{figure:structure} is the layer structure of the classical processing for topological 
error correction.  In total there are four stages, parity calculation, tree creation, 
minimum weight matching, and the quantum controller.  
As this analysis is only considering the classical requirements for base level error correction 
processing, we will not discuss the structure for the quantum controller.  This top-level processor 
will be responsible for the application of active quantum algorithms on the topological lattice, 
given inputs from both a quantum algorithm compiler, and error data outputted from the 
correction processors.  In the future, we will be examining the requirements for this top level 
processor, but for now we omit details regarding its structure and operation.

The error correction in this model requires identifying all  
cells which have changed parity 
and reliably identifying pairs of parity changes associated with a physical error chain.  
In order for error 
correction to be effective, we assume a standard, stochastic, qubit error model.  As all 
standard error channels (all errors except qubit loss) can be mapped under the topological 
model to phase errors~\cite{RHG07,FG08}, we can, without loss of generality, assume that each 
qubit experiences a phase error, $\sigma_z \equiv Z$, with probability $p$.  Therefore, a stochastic 
hierarchy exists with respect to longer error chains.  

As the probability of a single phase error is given by $p < 1$, the probability of an 
$d$ error chain is $O(p^d) \ll 1$.  Therefore, for a given set of 
parity results, 
the most likely set of physical errors producing the classical measurement pattern is the 
set of end point pairings where the total length of all connections is minimized.  

Classical results stemming from the detection layer are used to calculate the parity for 
all unit cells for some total volume.  The co-ordinates of all cells which have experienced 
a parity flip are stored in a classical data structure.  Minimum weight 
matching algorithms~\cite{CL65,E67,CR99,K09} are then used to calculate likely error chains corresponding to the detected endpoints.  Once chains are calculated, the Pauli frame (the current 
eigenvalue condition for all cells relative to their initial state)
of the computation, within the quantum controller, is updated with the new error information.  

\begin{figure}[ht]
\begin{center}
\resizebox{80mm}{!}{\includegraphics{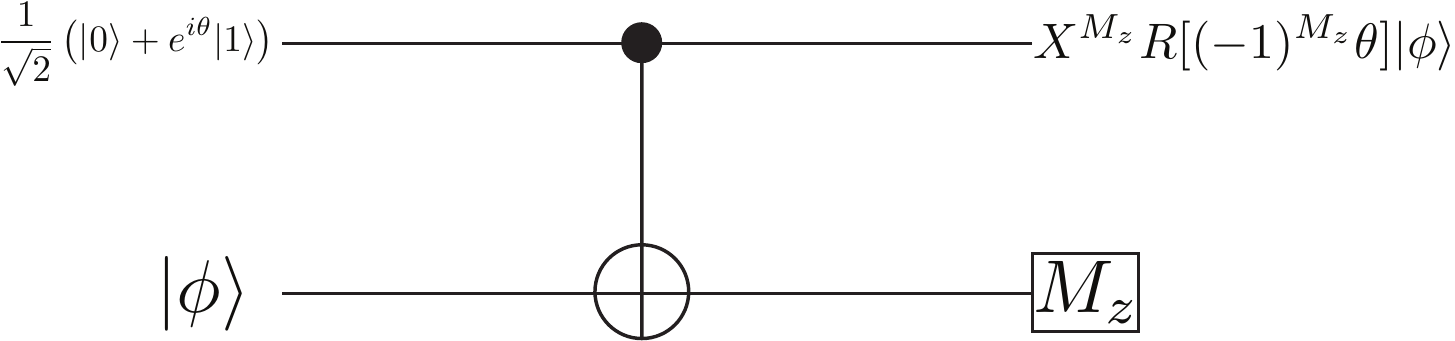}}
\end{center}
\vspace*{-10pt}
\caption{Standard teleportation 
circuit required to perform the rotation $R_z(\theta)$ on an arbitrary 
quantum state via the preparation of an appropriate ancilla state. The presence of 
a bit-flip $\equiv X$ errors on the qubit affects the gate.   
Error information during quantum processing must be up to date upon applying 
these types of gates in order to ensure rotations are applied in the correct direction.}
\label{figure:Tgate}
\end{figure}

The frequency of error correction in the lattice is dictated by the application of the non-Clifford 
gate
\begin{equation}
T = \begin{pmatrix}
1 & 0 \\ 0 & e^{\frac{i\pi}{4}}
\end{pmatrix}.
\end{equation}
This gate is required to generate a universal set of operations~\cite{NC00}, and in 
the topological model is 
achieved via the injection of multiple, low fidelity, singular ancilla, magic-state distillation~\cite{BK05+,R05} and 
the application of teleportation protocols with information qubits.  In order to successfully apply 
logical $T$ gates, error information must be obtained for the qubit undergoing the teleported $T$ 
gate prior to application of the correction.  Fig.~\ref{figure:Tgate} illustrates the teleportation 
protocol to implement an $R_z(\theta)$ rotation.  If a logical $X$ 
error exists on the state prior to teleportation, the condition
\begin{equation}
R_z(\theta)X\ket{\phi} = XR_z(-\theta)\ket{\phi}
\end{equation}
implies that this error must be known before teleporting the rotation $R_z(\theta)$.  If the 
error is detected after teleportation, the conjugate rotation $R_z(-\theta)$ will actually be 
applied.
Therefore, the classical processing of the minimum weight matching algorithm will have to 
occur at a comparable rate to the logical gate rate of the preparation network to 
ensure up-to-date error information for all logical qubits is available when teleported gates 
are applied.  

As detailed in Ref.~\cite{DGOH07}, the clock cycle of the preparation network 
can vary from nanoseconds to microseconds, depending on 
the system utilized to construct the photonic chip.  Hence our goal in this investigation is 
to determine, for a given failure probability of the {\em quantum} component of the computer, 
how quickly the network be operated such that all classical processing can be performed 
utilizing today's technology.  

\section{Layers two and three: Minimum weight matching}
\label{sec:benchmarking}

Calculating the minimum weight matching of the classical parity data stemming from the 
first layer of the classical processing network is the essential requirement for  
error correction in the topological model.  The parity processing layer is designed to simply output 
co-ordinate tuples for all parity changes in the lattice to this next layer.  The relevant question is, 
can the minimum weight processing of this data be performed over a large volume of the 
cluster in a comparable time frame to the quantum preparation network?

\subsection{Minimum Weight Matching benchmarking}

Classical algorithms for determining the minimum weight matching of a connected 
graph are well known with a run-time polynomial in the total number of nodes~\cite{CR99,K09}.  Such 
algorithms are derived from the original Edmonds' Blossom algorithm~\cite{E67} and 
for our benchmarking tests we have used 
Blossom V~\cite{K09}.  However, due to the nature of our problem, there are several adaptations 
that can be made to optimize the algorithm further.  

Typical minimum weight matching algorithms accept a list of $N$ 
co-ordinates, $N-even$, and of weighted edges (in this case, 
lattice separation between nodes) such that the corresponding graph is 
completely 
connected.  The output is then a list of edges such that every node is touched by exactly 
one edge and the total weight of the edges is a minimum.  
For the purposes of TCQC we can optimize 
by considering the specifics of the problem.  

\begin{figure}[ht]
\begin{center}
\resizebox{80mm}{!}{\includegraphics{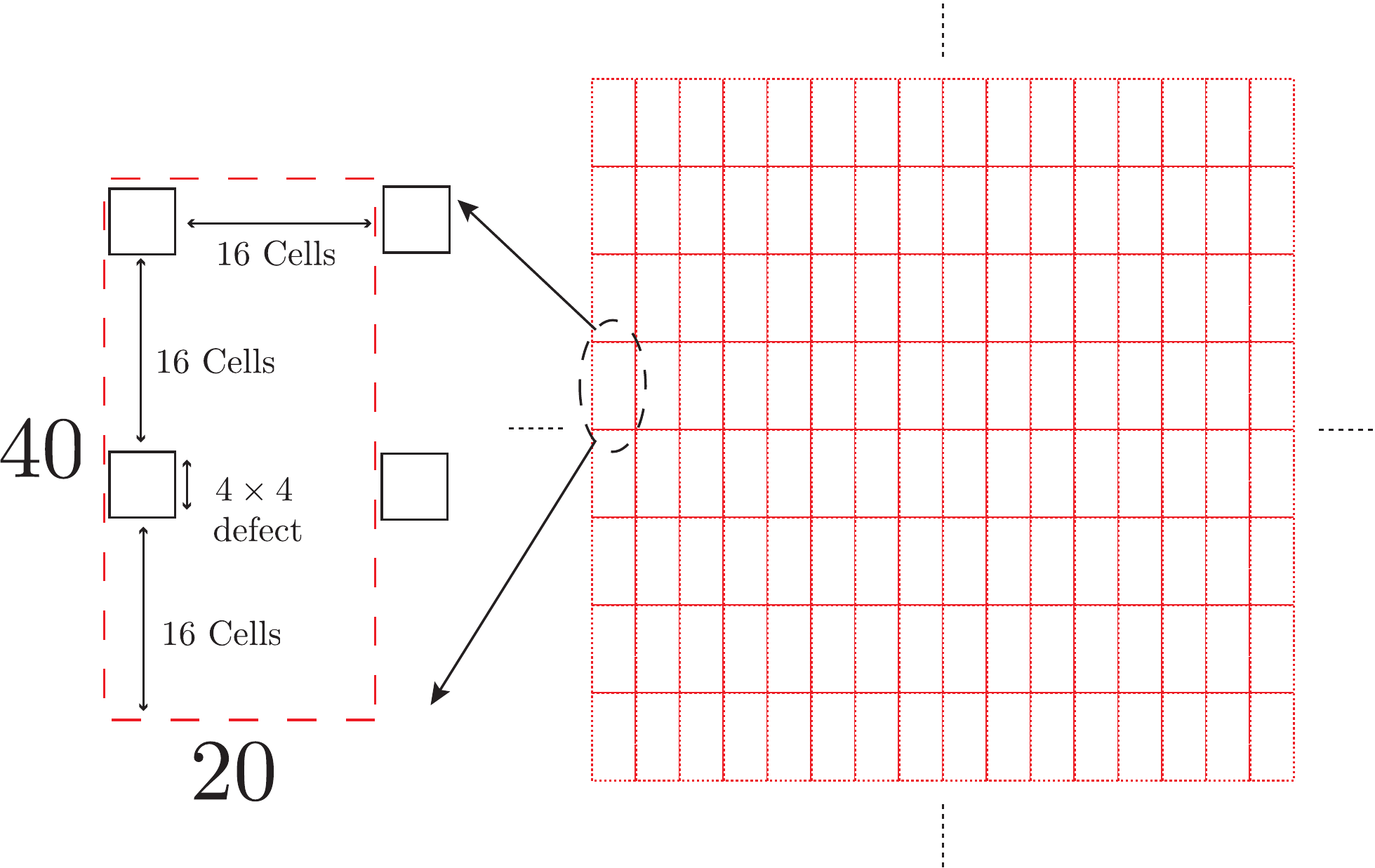}}
\end{center}
\vspace*{-10pt}
\caption{(Taken from Ref.~\cite{DFSG08}).  
Cross section of a large topological lattice.  Qubits are defined within a cluster region of 
approximately $40\times 20$ cells.  The actual qubit information is stored in pairs of defect regions 
(artificially created holes in the lattice).  As undetectable error chains occur when errors connect 
two boundary regions, topological protection is achieved by keeping defect qubits well separated 
from boundaries and from each other.}
\label{figure:computer}
\end{figure}

Due to the stochastic hierarchy of errors in the qubit model, and the assumption that the operational 
error rate of the computer is low, $p \ll 1$, the most likely patterns of errors are simply sets 
of sparse single errors causing two adjacent cells to flip parity.  The probability of 
obtaining longer and longer errors chains is increasingly unlikely.  
\begin{figure}[ht!]
\begin{center}
\resizebox{85mm}{!}{\includegraphics{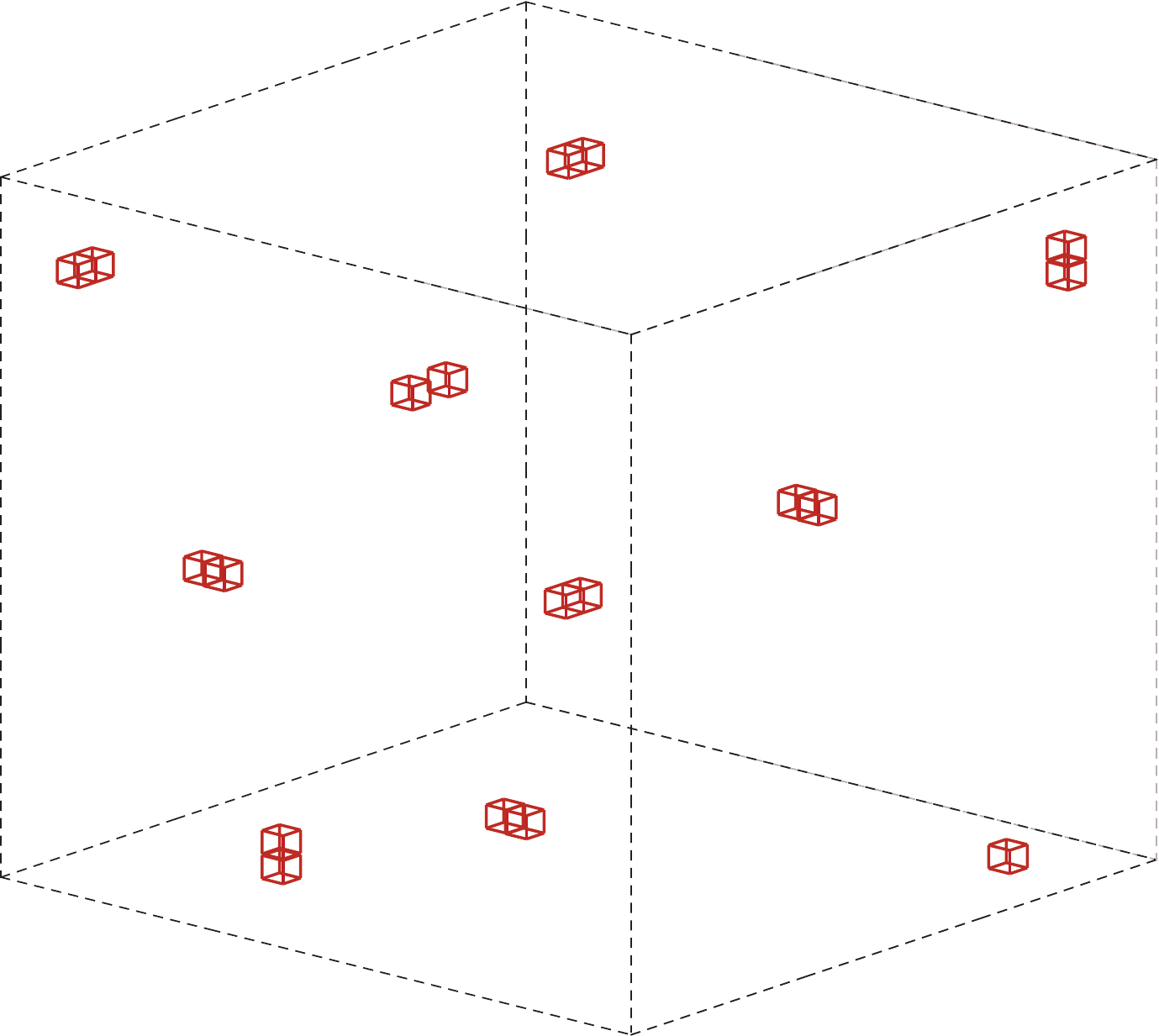}}
\end{center}
\vspace*{-10pt}
\caption{Likely structure of the error data for a large volume of cluster cells.  As we are assuming 
the operational error rate of the quantum computer is low, $p\ll 1$, long error chains become 
increasingly unlikely.  Hence, cell parity flips will most likely be sparse and, generally, pairs of 
parity flips will be isolated.  This property of the computational model allows a certain amount 
of optimization of the classical requirements for minimum weight matching.}
\label{figure:errors}
\end{figure}
Therefore, 
for a given volume of classical parity results, erred cells will tend to be clustered into small 
groups, as illustrated in Fig.~\ref{figure:errors}.  Additionally, we can consider the computational 
lattice structure.  Fig.~\ref{figure:computer}, taken from Ref.~\cite{DFSG08}, illustrates 
the 2D cross-section of the cell structure once logical qubits are defined.  
In this example, defects are separated from each 
other (and from the edge of the lattice) by a total of 16 cells.  
Logical errors occur when error chains connect 
two boundaries in the cluster.  If an error chain spans more than 8 cells, the correction inferred from the endpoints is likely to be incorrect, resulting in defects connected by a chain of errors - a logical error.  

This allows us to set a maximum edge length that is allowed between connections in the minimum 
weight matching algorithm.  Instead of creating a completely 
connected graph structure from the classical 
data, we instead create multiple smaller subgraphs, with each subgraph having no connections 
of weight greater than a maximum edge parameter $m_e$.   
As the separation and circumference of defects within the lattice determines the effective 
distance of the quantum code $m_e$ can safely be chosen such that $m_e = \lfloor d/2\rfloor$.  This 
approximation ensures that {\em all} error chains throughout the lattice with a weight 
$\leq \lfloor d/2\rfloor$ are connected within the classical processing layer.  This 
classical approximation is defined to fail if a single length 
$ > \lfloor d/2 \rfloor$ error chain occurs during computation.  This 
is much less likely than failure of the code itself which can occur as the result of 
numerous, not necessarily connected, arrangements of $\lfloor (d-1)/2 \rfloor$ 
errors.  

It should be noted that this approximation does not neglect {\em all} error chains longer than
$m_e$.   However, the classical error correction 
data has no knowledge regarding the actual path an error chain has taken through 
the lattice.  The $m_e$ approximation neglects all error chains with {\em endpoints} 
separated by $> m_e$ cells.  

By making this approximation, we speed up the classical processing for minimum 
weight as, on average, the algorithm will be run on a very sparse graph structure. 

\subsection{Classical simulations as a function of total cluster volume and code-distance.}

Given the above classical approximation, we have benchmarked the 
Blossom V algorithm~\cite{K09} as a function of the total volume of the cluster for various 
distances of the quantum code, $d$ and hence $m_e$.  

In order to investigate the classical processing requirements of the system, we will assume a 
physical error rate of the quantum computer.  As with previous 
investigations into this system we will be assuming, throughout this discussion, that the 
quantum computer operates at a physical error rate of $p=10^{-4}$~\cite{DFSG08,DMN08} and the fault-tolerant 
threshold is $p_{th} \approx 0.61\%$~\cite{RHG07}.  
\begin{figure*}[t]
\begin{center}
\resizebox{180mm}{!}{\includegraphics{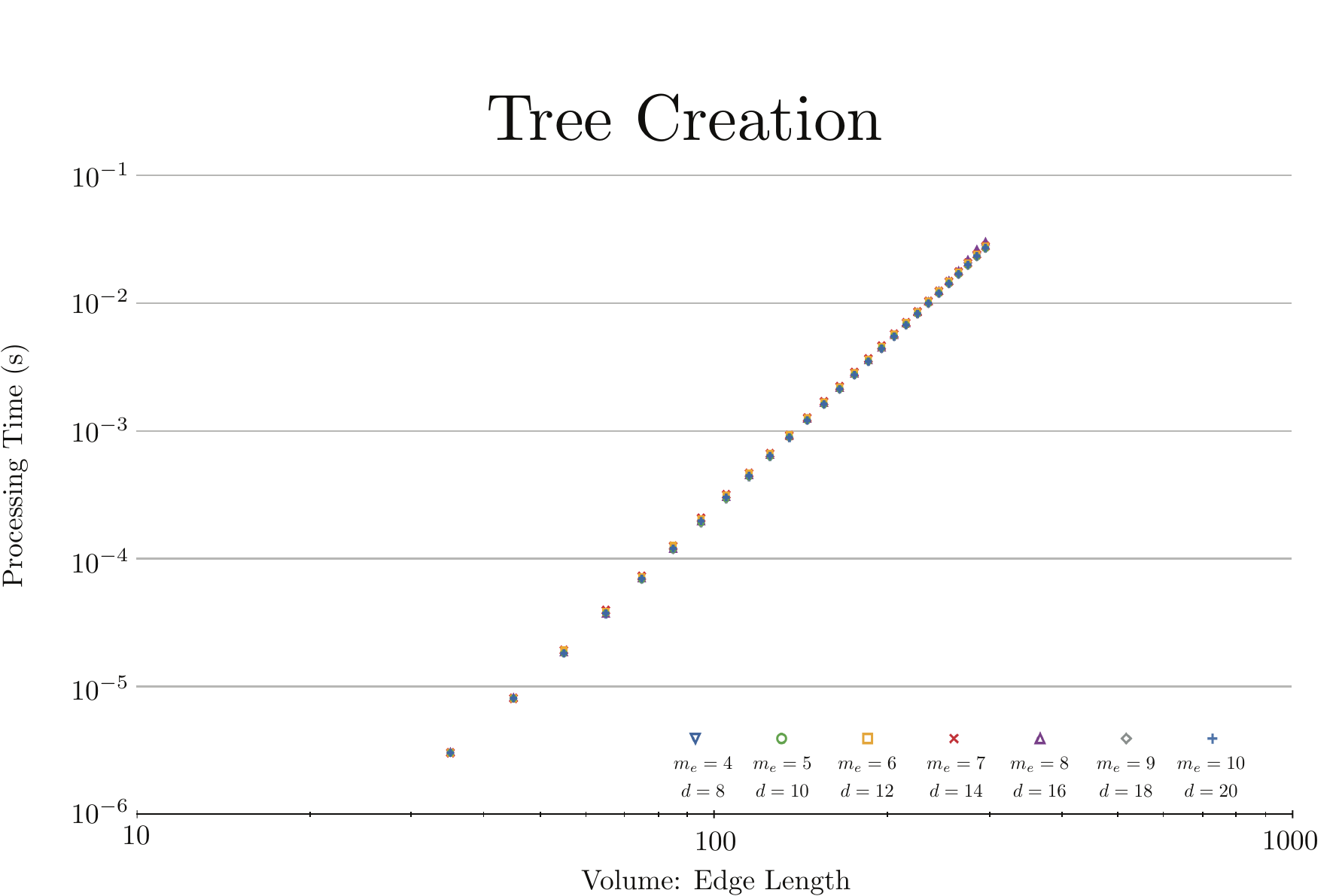}}
\end{center}
\vspace*{-10pt}
\caption{Benchmarking data for the tuple $\rightarrow$ tree creation processing layer.  
Taken with $10^4$ statistical 
runs, an operational error rate of 
$p=10^{-4}$ for various sized codes, $d = (8,10,12,14,16,18,20)$, $m_e = [4,..,10]$.  
Notice that $m_e$ does not significantly alter the processing time.}
\label{figure:plot1}
\end{figure*}
\begin{figure*}[ht!]
\begin{center}
\resizebox{180mm}{!}{\includegraphics{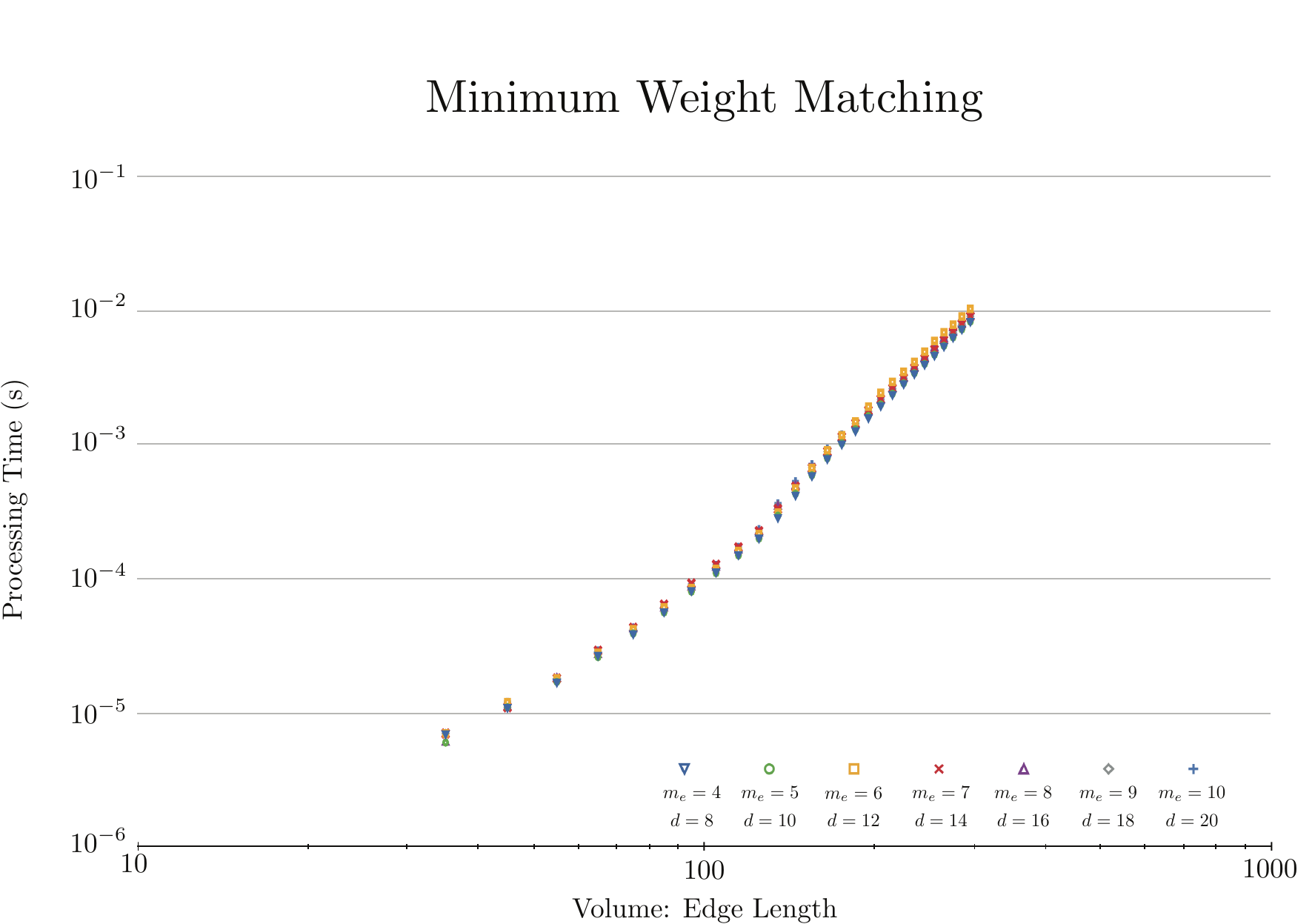}}
\end{center}
\vspace*{-10pt}
\caption{Benchmarking data for the minimum weight matching processing layer.  The 
simulation conditions for this data set 
are identical to the tree creation layer shown in Fig.~\ref{figure:plot1}.}
\label{figure:plot1a}
\end{figure*}
Given this base assumption, Figs.~\ref{figure:plot1} and~\ref{figure:plot1a} 
examine the processing time of the modified Blossom V algorithm run 
on a MacBook Pro (technical details of this computer are summarized in 
Appendix.~\ref{sec:appA}).  For various volume sizes, $V$, 
from $35^3$ to $295^3$ cells, random single qubit $Z$ errors were generated with a 
probability of $10^{-4}$.  The processing time for each value of $V$ was examined 
for different distance quantum codes and hence different values of the 
approximation parameter, $m_e = d/2$.  
A list of cell co-ordinate tuples, $(i,j,T)$, corresponding to the endpoints of error chains 
(cells with changed parity) was constructed.  This list is the input to the classical processor as, in 
practice, it would be provided directly by the hardware.  

The tuple data was then processed in two stages.  In the first stage, which we denote as the tree 
creation layer, an 8-way sort/search tree was created from the tuple information and then 
used to generate a list of connections (edges) between cells and 
their distances (weights) [Fig.~\ref{figure:plot1}].  
The second stage of the simulation applied the 
Blossom V minimum weight matching algorithm to the generated sparse graph structures 
[Fig.~\ref{figure:plot1a}].  
The benchmarking data was taken using $10^4$ runs per data point. 

For each cluster volume, code distances $d=[8,10,12,14,16,18,20]$, corresponding 
to $m_e = [4,..,10]$ were simulated.  
For both the tuple $\rightarrow$ tree creation and the minimum weight matching, Figs.~\ref{figure:plot1} 
and~\ref{figure:plot1a} illustrate that the approximation parameter, $m_e$, does not significantly 
alter the total simulation time and that for a given ($V,m_e$), 
 tree creation and minimum weight matching take a similar amount of time.



\subsection{Parallelizing the algorithm}

The numerical simulations shown in Figs.~\ref{figure:plot1} and~\ref{figure:plot1a} 
clearly illustrate that the 
minimum weight matching subroutine cannot be run over the entire lattice used for 
TCQC.  As a rough estimate, a mainframe device such as the one 
introduced in Ref.~\cite{DMN08} consists of a lattice cross section 
measuring $(5\times 10^5)\times (4\times 10^3)$ 
unit cells.  Clearly in order to achieve classical processing speeds of the 
order of microseconds (for any distance topological code), 
either the classical fabrication of the processing 
equipment must allow for at least a 10--15 order of magnitude speed up from a standard Laptop 
or the application of the tree creation and minimum weight subroutines must be highly parallelized. 

\begin{figure*}[ht!]
\begin{center}
\resizebox{180mm}{!}{\includegraphics{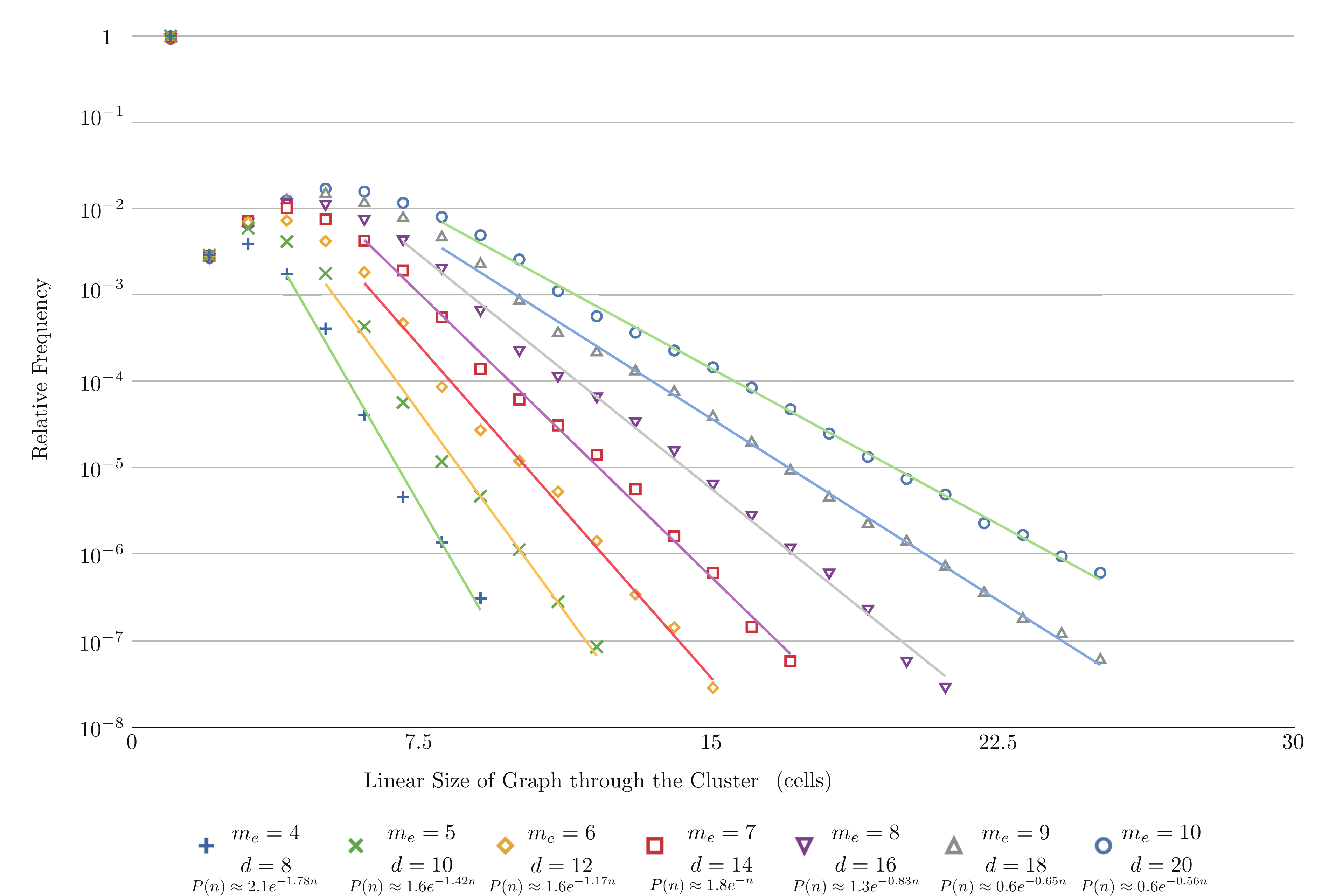}}
\end{center}
\vspace*{-10pt}
\caption{Volume independent distribution of connected component sizes for 
cluster error data.  Shown above is the distribution of the maximum {\em linear} 
size of each connected component for error data, simulated 
with $p=10^{-4}$, $m_e = [4,..,10]$ and 
performed with $2\times10^6$ statistical runs (giving a total number of 
connected components $\approx 3\times 10^7$).  
Simulations were performed using 
the Floyd-Warshall algorithm~\cite{F62,W62}, but instead of calculating the maximum 
distance between any two nodes in a connected graph we instead calculate the 
maximum physical distance through the cluster (in a single spatial dimension) 
between any two connected 
nodes.  These results allow us to estimate the edge length of a cube of 
sufficient size to encapsulate {\em all} connected components at a given 
value of $m_e$.  Performing an approximate exponential fit to 
the decay of these curves allow for the estimation of the probability of obtaining 
very large connected components.  
Appendix~\ref{sec:appB} presents 
further simulation results explaining the general properties of this curve.
}
\label{figure:peaks}
\end{figure*}

Due to the approximation made to the Blossom V algorithm, parallelizing the classical 
processing is possible.  The $m_e$ approximation to the subroutine prohibits the 
establishment of graph connections between two cluster cell co-ordinates separated by a 
distance $> m_e = d/2$.  In Fig.~\ref{figure:peaks} we illustrate the 
relative frequency of different sized connected components within the lattice at an 
error probability of $p=10^{-4}$, for $m_e = [4,..,10]$.  
These simulations were performed using the Floyd-Warshall algorithm~\cite{F62,W62} 
obtained for a volume region of $V=50^3$, with $2\times 10^6$ statistical runs (resulting 
in approximately $3\times 10^7$ connected components in total).   
In these simulations, the size of each connected component in the lattice, $n(m_e)$, 
does {\em not} 
represent the longest path through the graph.  Instead it represents the physical edge 
length through the cluster of a cube of sufficient size to fully contain 
each connected component.  In Appendix~\ref{sec:appB} we provide additional simulations showing the distribution of 
connected components within the cluster to assist in the explanation of Fig.~\ref{figure:peaks}. 

\begin{figure*}[ht!]
\begin{center}
\resizebox{180mm}{!}{\includegraphics{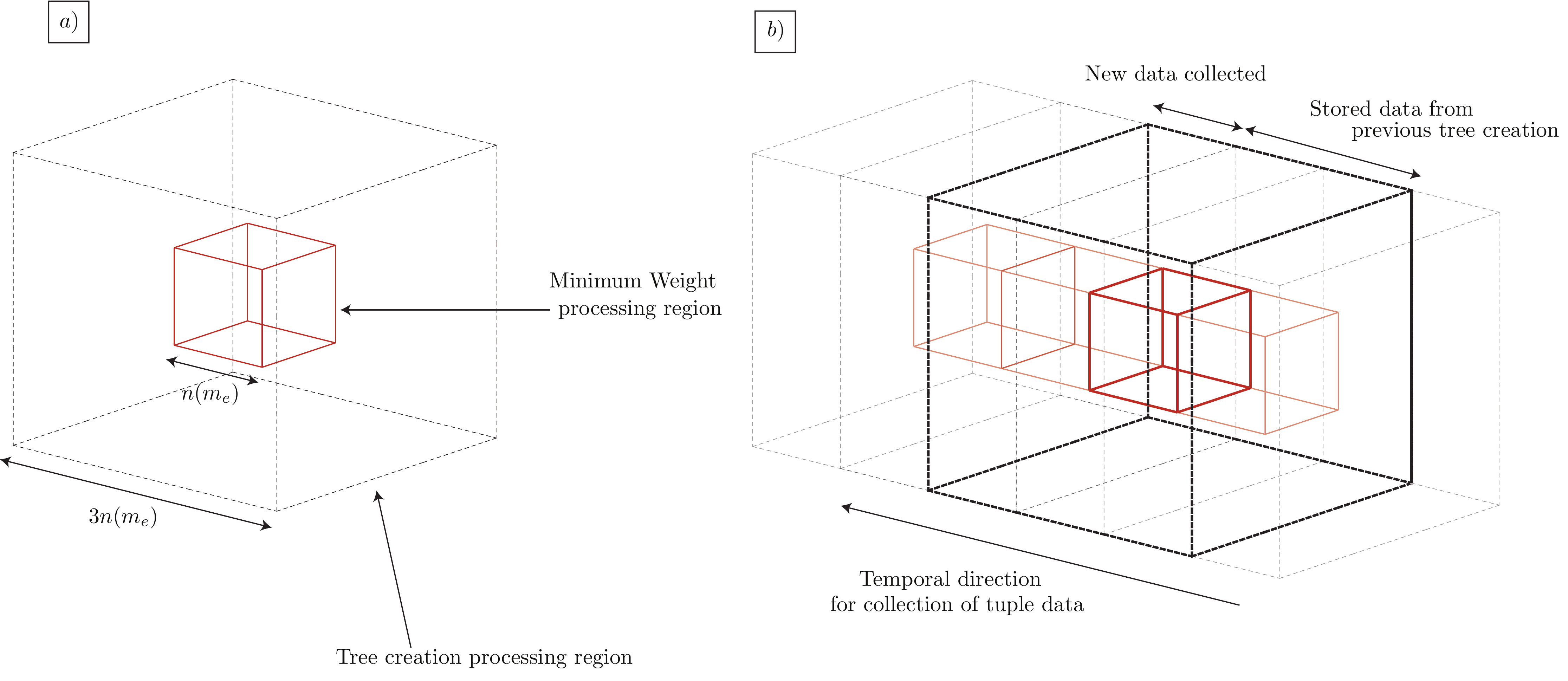}}
\end{center}
\vspace*{-10pt}
\caption{Tree creation structure and minimum weight matching processing structure for 
parallel application of minimum weight matching.  a) illustrates the volume regions for 
the two classical processes.  For a given approximation parameter, $m_e$, a volume of 
$27n(m_e)^3$ of tuple data is sent to an independent tree creation process.  Once processed, 
any subgraph with at least one vertex in the inner volume region of $n(m_e)^3$ is sent to 
a separate instance of minimum weight processing for this region.  Parallelization is achieved 
by interlacing the outer volumes such that the inner volumes touch.  b) illustrates the 
temporal processing of this data, where a given tree creation volume is processed through 
time.  For each instance of the tree creation process, 67\% of the tuple data is taken 
from the previous instance of tree creation (as the previous volume overlaps the new 
volume by two-thirds).  The remaining 33\% of the data is collected directly from the measurement 
of photons.  Therefore both the tree creation process and minimum weight process must 
be completed within the time for collection of the new data in order to keep up with 
the quantum clock speed.}
\label{figure:maxchain}
\end{figure*}

Parallelizing the minimum-weight matching procedure requires subdividing up a large volume 
of the cluster into smaller regions such that each instance of the tuple $\rightarrow$ 
tree creation and the minimum weight algorithms faithfully return the same results as 
processing the entire volume (up to the failure probability of the computer).  
In Fig.~\ref{figure:peaks} we provide an approximate scaling of the decay of each 
curve, representing the volume independent relative frequency of connected 
component with a linear size, $n(m_e)$, in the cluster.  In order to parallelize classical processing, 
two regions are defined.  Fig.~\ref{figure:maxchain}a. illustrates.  

The inner volume defines the minimum weight processing region while the outer volume, 
with an edge length of $3\times$ the inner volume, defines the tree creation processing 
region.  During tuple $\rightarrow$ tree creation, if any connected component contains 
at least {\em one} vertex within the inner volume it is sent to independent instances of minimum 
weight matching.  Provided that the edge length of these regions are large enough, then 
{\em any} connected component with at least one vertex in the inner volume will be 
fully contained within the outer volume with high probability.  

To determine the size of these processing regions we utilize the decay of the curves in 
Fig.~\ref{figure:peaks}.  The probability of failure when parallelizing classical processing 
should be approximately the same as the failure probability of the quantum computer itself.  
In order to determine these failure probabilities, we consider the volume of the cluster required 
to perform a logical CNOT operation as a function of $m_e$.  Fig.~\ref{figure:computer} 
illustrates the logical structure of the lattice.  Each logical qubit cell in the cluster consists of a 
cluster cross-section measuring $(2d+d/2)(d+d/4) = 25m_e^2/2$ cells.  A CNOT gate 
requires 4 logical cells and the depth through the cluster required to perform the gate is 
$(2d+d/2) = 5m_e$ cells~\cite{RHG07}.  Hence the total cluster volume for a CNOT operation is 
$V = 250m_e^3$ cells.

The failure probability of the quantum computer during a {\em logical} CNOT is approximately,
\begin{equation}
p_L(m_e) \approx 1 - (1- {\Omega(m_e)})^{\lambda(m_e)}
\label{eq:prob}
\end{equation}
where,
 \begin{equation}
 \Omega(m_e) \approx \left(\frac{p}{p_{th}}\right)^{d/2} = 10^{-2m_e} 
 \end{equation}
is the probability of failure for a single logical qubit a single layer thick, with a 
fault-tolerant threshold of 
$p_{th} \approx 0.61\%$, $p=10^{-4}$ and 
$\lambda(m_e) = 4(2d+d/2) = 20m_e$ is the number of such layers of the cluster 
that need to be consumed to perform a logical CNOT operation.

Given the failure rate of the quantum computer, we utilize the data from Fig.~\ref{figure:peaks} 
to determine the edge length of a volume large enough to encapsulate all connected 
components with a probability approximately equal to Eq.~\ref{eq:prob}.  As Fig.~\ref{figure:peaks} 
represent relative frequencies (the probability of a connected component of size $n(m_e)$, 
relative to a connected component of size one) we scale $P(n,m_e)$ by the number of 
isolated errors expected in a cluster volume required for a CNOT.  Hence,
\begin{equation}
6p\times 250m_e^3P(n,m_e) = 1 - (1- {\Omega(m_e)})^{\lambda(m_e)}
\label{eq:prob2}
\end{equation}
where the factor of 6 comes from the 6 independent qubits per unit cell of the cluster.  
Eq.~\ref{eq:prob2} is then solved for $n(m_e)$ giving,
\begin{equation}
n(m_e) = -\frac{1}{\beta(m_e)}\ln\left( \frac{1-(1-10^{-2m_e})^{20m_e}}{0.15\alpha(m_e)}\right)
\end{equation}
where $[\alpha(m_e),\beta(m_e)]$ are the scaling parameters shown in Fig.~\ref{figure:peaks}.

The values of $n(m_e)$ for $m_e = [4,..10]$ and the probabilities of 
the logical CNOT failure and equally the probability of a connected component 
of size greater than $n$ are shown in Tab.~\ref{tab:n}.  

\begin{table}[ht!]
\begin{center}
\vspace*{4pt}   
\begin{tabular}{c|c|c|c}
$m_e$ & CNOT failure & $n(m_e)$ & $P(\text{connected component } > n)$\\
\hline
4 & $O(10^{-7})$ & 10 & $O(10^{-7})$\\
5 & $O(10^{-8})$ & 15 & $O(10^{-8})$\\
6 & $O(10^{-10})$ & 23 & $O(10^{-10})$\\
7 & $O(10^{-12})$ & 32 & $O(10^{-12})$\\
8 & $O(10^{-14})$ & 44 & $O(10^{-14})$\\
9 & $O(10^{-16})$ & 62 & $O(10^{-16})$\\
10 & $O(10^{-18})$ & 81 & $O(10^{-18})$\\
\end{tabular}
\caption{Maximum edge length, $n(m_e)$, of a cube of sufficient size in the lattice 
to encapsulate all connected components of the tuple $\rightarrow$ tree creation 
graph structure.  
The last column is the probability that a connected component of the graph is unbounded 
by a cube of volume $n^3$ and Eq.~\ref{eq:prob2} ensures that this occurs with approximately 
the same probability as the CNOT failure rate of the topological computer.}
\label{tab:n}
\end{center}
\end{table} 

In Tab.~\ref{tab:n} we give the size of the processing regions 
for parallelizing both the tuple $\rightarrow$ tree creation and minimum weight matching 
processes such that the probability of any connected component within the inner 
region unbounded by the boundary of the outer region is approximately the same as 
the failure rate of the quantum computer.  The value $n(m_e)$ therefore sets the edge length 
of the inner and outer volume regions.

Using this estimate, the tree creation layer becomes an interlaced network, 
with each individual instance of tree 
creation operating over a volume of $V\approx 27n(m_e)^3$.
As we can neglect larger connected components 
(which occur with probability roughly equal to the 
probability of quantum failure), 
any connected component that contains at least one vertex 
within a central volume $V\approx n(m_e)^3$ will be fully contained within the outer 
volume of $V \approx 27n(m_e)^3$.  
The central volume represents the region that is sent to separate instances of the 
minimum weight matching algorithm.  The tree creation layer is interlaced such that each of the 
central volumes touch but do not overlap.   

This processing structure 
ensures that multiple parallel instances of minimum weight matching will produce identical results 
to an individual instance of minimum weight matching run over the entire volume.  As the tree creation 
layers are interlaced, tree structures that cross the boundaries of two inner volumes regions will 
be sent to two independent instances of minimum weight matching.  After processing, 
these duplicate results will simply be removed from the final error list.  

We can now combine the results from Tab.~\ref{tab:n} with the simulation data of 
Figs.~\ref{figure:plot1} and~\ref{figure:plot1a} to determine the maximum size and speed of a 
quantum computer such that error correction data can be processed sufficiently quickly.
Fig.~\ref{figure:maxchain}b.  illustrates how error correction data is collected as qubits along the 
third axis of the cluster are sequentially measured.  As the tuple $\rightarrow$ tree creation 
processing layer consists of an interlaced set of $V=27n(m_e)^3$ cubes, two thirds of 
the data for any given volume is taken from the previously collected results while the 
final one third consists of newly collected data.  Therefore the processing ``window" 
available for each instance of the tuple $\rightarrow$ tree creation (and hence 
each instance of minimum weight matching associated with each tree creation volume) 
is the time required to collect this new data.  

The optical network illustrated in Sec.~\ref{sec:flow} has each parity calculation 
from the detector banks, for each unit cell, occurring 
over three successive cluster faces and, in the absence of loss, occur every 
2$T$, where $T$ is the separation period of photons (the quantum clock rate).
Tree creation from the parity tuples for a given volume element utilizes the last 
$9n(m_e)^2\times 2 n(m_e)$ 
tuple information from the previous instance of tree creation and must 
store the same amount for use in the next tree creation subroutine.  
Parity tuple $\rightarrow$ tree creation processing is repeated for each $27n(m_e)^3$ volume
every $2Tn(m_e)$ seconds.  

Taking the simulation data from Fig.~\ref{figure:plot1} (as tuple $\rightarrow$ tree-creation 
is the slower of the two processes), the fastest clock 
cycle, $T_{\text{min}},$ can be 
calculated for each value of $m_e$ as,
\begin{equation}
T_{\text{min}}(m_e) = \frac{t(3n(m_e))}{2n(m_e)}.
\end{equation}
where $t(3n(m_e))$ is the processing time as a function of the edge length of the 
processing volume, $3n(m_e)$, shown in Fig.~\ref{figure:plot1a}.  The 
results are shown in Tab.~\ref{tab:n2}. 
\begin{table*}[ht!]
\begin{center}
\vspace*{4pt}   
\begin{tabular}{c|c|c|c|c|c|c}
$m_e$ & CNOT failure & $n(m_e)$ & $T_{\text{min}}(m_e) (\mu s)$ & ``window", $2n(m_e) (\mu s)$ & CNOT operating Freq. & Processing instances / 
Logical qubit, (I)\\
\hline
4 & $O(10^{-7})$ & 10 & 0.06 & 19 & 105 kHz & 8.7\\
5 & $O(10^{-8})$ & 15 & 0.28 & 31 &  18 kHz & 5.3\\
6 & $O(10^{-10})$ & 23 & 1.1 & 46 &  4 kHz & 3.4\\
7 & $O(10^{-12})$ & 32 & 3.2 & 64 & 1 kHz & 2.4\\
8 & $O(10^{-14})$ & 44 & 9.3 & 87 & 0.3 kHz & 1.7\\
9 & $O(10^{-16})$ & 62 & 28 & 124 & 0.1 kHz &1.0\\
10 & $O(10^{-18})$ & 81 & 68 & 162 & 37 Hz & 0.77\\
\end{tabular}
\caption{Maximum size and speeds for topological quantum computers when classical processing is performed utilizing the benchmarking data of Figs.~\ref{figure:plot1} and~\ref{figure:plot1a}.  
The failure of logical CNOT gates defines the size of the computer, $\approx 1/KQ$, where 
$Q$ is the number of logical qubits in the system and $K$ is the total number of logical 
time steps available for an algorithm.  $T_{\text{min}}(m_e)$ defines the maximum speed the 
quantum network can be operated such that error correction data can be processed 
sufficiently quickly.  The processing ``window", independent of the benchmarking data, 
is related to the parallelization of classical processing.  Processing instances / Logical qubit 
defines how many classical processes are required for a lattice cross section housing one 
logical qubit.}
\label{tab:n2}
\end{center}
\end{table*} 

The last column in Tab.~\ref{tab:n2} calculates the total number of processing instances 
required per logical qubit in the computer.  This is calculated as the ratio of the 
cross-sectional area of a logical qubit to the cross-sectional area of the minimum weight 
matching processing volume. 
\begin{equation}
 I = 4\times \frac{(2d+d/2)(d+d/4)}{n(m_e)^2} = 4\times \frac{25m_e^2}{4n(m_e)^2}
 \end{equation}
 the factor of 4 is introduced since each instance of minimum weight matching has an 
 associated tuple $\rightarrow$ tree creation process and that primal and dual lattice 
 error correction is performed independently giving another factor of two. As the 
 size of the quantum computer increases (increasing $m_e$) this ratio decreases 
 as the scaling of $n(m_e)$ is approximately $n(m_e) \approx O(m_e^2)$.

While the total size and speed of a topological quantum computer will ultimately 
be governed by the experimental accuracy in constructing each quantum component, the 
results shown in Tab.~\ref{tab:n2} are promising.  Assuming that quantum fabrication can 
reach and accuracy of $p=10^{-4}$, current classical technology is quite sufficient to 
process error correction data for a large range of computer sizes.  The logical failure rate of 
the CNOT gate approximately defines the size of the computer, $p_L(\text{CNOT}) \approx 
O(1/KQ)$, where $Q$ is the number of logical qubits in the computer and $K$ is the number 
of logical time steps in a desired quantum algorithm (note that the application of 
non-clifford gates will lower this effective size further).  Even for a small 
topological computer ($m_e=4$ has sufficient protection for 1000 logical qubits running 
an algorithm requiring approximately $10^4$ time steps) less than ten classical processing 
instances are required per logical qubit, a quantum network run at $\approx 17$MHz with 
a logical CNOT operating frequency of $\approx 100$kHz.

The classical processing power utilized in this investigation is clearly not specially designed for 
the task of operating a topological computer.  Not only can we safely assume that classical 
processing power will increase before the advent of a large topological computer, but the 
design and implementation of both specialized hardware and more optimal coding should also 
result in significant increases in the achievable operational frequency of the quantum network 
and logical gates.  
More recent analysis has suggested that possible operational frequency of the quantum 
network could reach the 
$100$MHz level~\cite{SGMNH08}.  In this case, if classical processing could result in a 2-3 
order of magnitude speed up when moving to optimized hardware and software, 
current {\em classical}
technology would be sufficient for a quantum computer capable of a logical error rate 
$\approx O(10^{-14})$ and a logical CNOT frequency of $\approx 0.3$MHz.  

\section{Observation and Conclusions}

In this work we have focused exclusively on the classical requirements to perform the underlying 
error correction processing for the network.  As the 
error correction procedures can be thought of as the base-level processing, the development of 
an appropriate quantum controller system is an obvious next step.   This higher level 
classical controller will essentially be responsible for the following,
\begin{enumerate}
\item The compilation of a user-designed quantum circuit into an appropriate measurement 
sequence on a topological cluster.
\item Direct control of quantum components within the measurement system of the topological 
cluster in order to change the measurement basis for the photon stream.
\item The dynamic allocation of cluster resources dependent on operational error rate.  Specifically, 
the fundamental partitioning of the lattice into appropriately separated defect regions for logical 
qubit storage.   
\item Accepting the data from the error correction processing layer to faithfully ensure 
accurate error correction is performed during computation.
\item Dynamical restructuring of the topological lattice partitioning to allow for ancilla preparation for 
non-Clifford quantum gates, and optimization of logical qubit/qubit interactions for specific quantum 
subroutines.
\end{enumerate}

This last point is one of the more interesting questions that can be addressed in this model.  
As we noted in the introduction, once the cluster lattice is prepared, data processing is 
performed via software.  The structure of the topological lattice essentially allows for 
qubit/qubit interactions in a 2D arrangement [Fig.~\ref{figure:computer}].  However, provided we 
have access to a large cluster lattice, we can envisage the dynamical creation of data 
pathways and ``flying defects" in order to speed up specific quantum subroutines.  
This could lead to some extremely interesting avenues of investigation in software control 
and optimization of a TCQC architecture.  

This analysis has demonstrated that the classical error correction requirements 
necessary to construct an optical quantum computer based on topological cluster states 
is certainly feasible given today's technology.   We have illustrated how minimum weight 
matching, required to process error data from the topological mainframe, can be 
optimized in such a way to allow for a massively parallelized processing network that 
can process information for large topological clusters. 

These results are very encouraging.  As with the quantum preparation network, the classical 
front end can also be constructed in a modular manner.  As the quantum preparation 
network is increased in sized via the addition of more photonic chips, the classical 
processing network is also expanded in a similar way.  Parity check processors, 
tuple $\rightarrow$ tree creation processors and minimum weight matching processors are 
also linked into a pre-existing network as its size expands.  
The results of this investigation give us a very optimistic outlook on the viability of the 
topological model as a possible avenue to achieve truly large scale quantum computation.

\section{Acknowledgments\protect}\label{Acknowledgments}

We would like to thank Rod Van Meter, Ashley Stephens, and Zac Evans for helpful discussions. 
We acknowledge the  support of MEXT, JST, HP and the EU project HIP and the support of the 
Australian Research Council, the Australian Government, and the US National Security Agency (NSA) 
and the Army Research Office (ARO) under contract number W911NF-08-1-0527.

\appendix
\section{Technical Specifications for simulations.}
\label{sec:appA}
The technical specifications to the computer used in benchmarking simulations are summarized 
below.

\begin{table}[ht!]
\begin{center}
\vspace*{4pt}   
\begin{tabular}{c|c}
Process & Benchmark\\
\hline
Floating Point Basic & 3.1Gflop/s  \\
vecLib FFT & 3.5 Gflop/s  \\
Memory Fill & 6.2 GB/s \\
\end{tabular}
\caption{Benchmarking for system processes for the MacBook Pro 2.2 GHz, 3GB RAM 
Benchmarking data was taken using the program XBench, version  1.3.}
\label{tab.comp}
\end{center}
\end{table} 

\section{Simulations of graph sizes for cluster errors.}
\label{sec:appB}

The simulations shown in Fig.~\ref{figure:peaks} illustrate the distribution of the largest 
physical distance through the cluster (in a single spatial dimension) 
between any two nodes for each connected component for graph structures established 
using the $m_e$ approximation detailed in the main text.  The following results illustrate 
the structure of this distribution in more detail.  
\begin{figure*}[ht!]
\begin{center}
\resizebox{150mm}{!}{\includegraphics{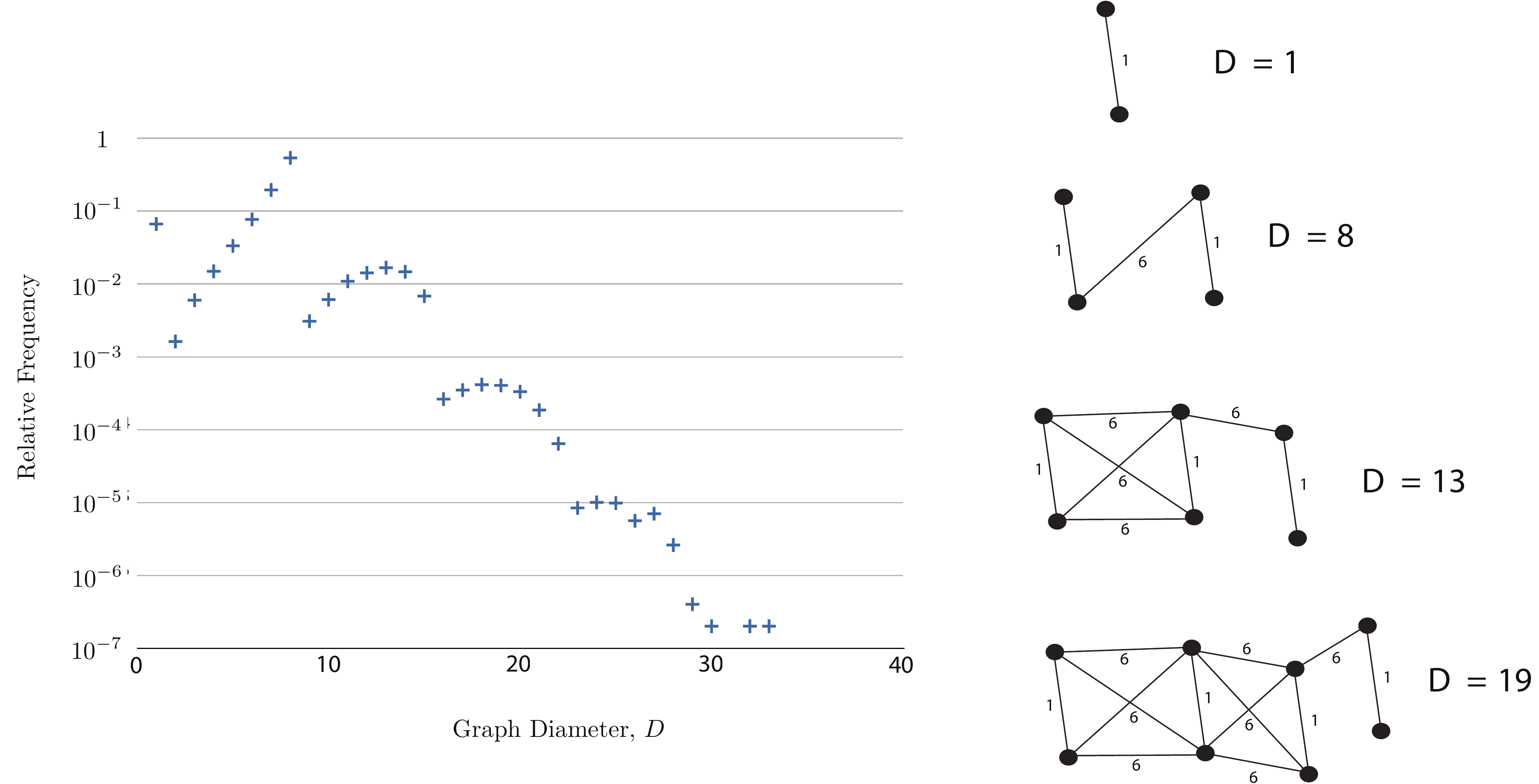}}
\end{center}
\vspace*{-10pt}
\caption{The maximum graph diameter over $10^5$ statistical runs 
utilizing $p=10^{-4}$, $m_e=6$ and a total cluster volume of $V=100^3$.  The Floyd-Warshall 
algorithm is designed to find the shortest pathway between any two connected nodes in the 
cluster and in these simulations we maximize over all possible connections.  Unlike 
Fig.~\ref{figure:peaks} we are examining the actual graph diameter, $D$, and {\em not} the 
linear separation of nodes in the physical cluster.  Here we are simply finding the {\em maximum} 
graph diameter in the complete data set, unlike Fig.~\ref{figure:peaks} which calculates the 
diameter of all connected components (hence these results exhibit volume dependence).  
At $p=10^{-4}$ 
and $m_e=6$, the graph structure for $D=8$ is the most probable.  Changing $p$ 
shifts which diameter graph is the most probable, while changing $m_e$ changes 
the values of $D$ where peaks occur.}
\label{figure:peaks2}
\end{figure*}

\begin{figure*}[ht!]
\begin{center}
\resizebox{150mm}{!}{\includegraphics{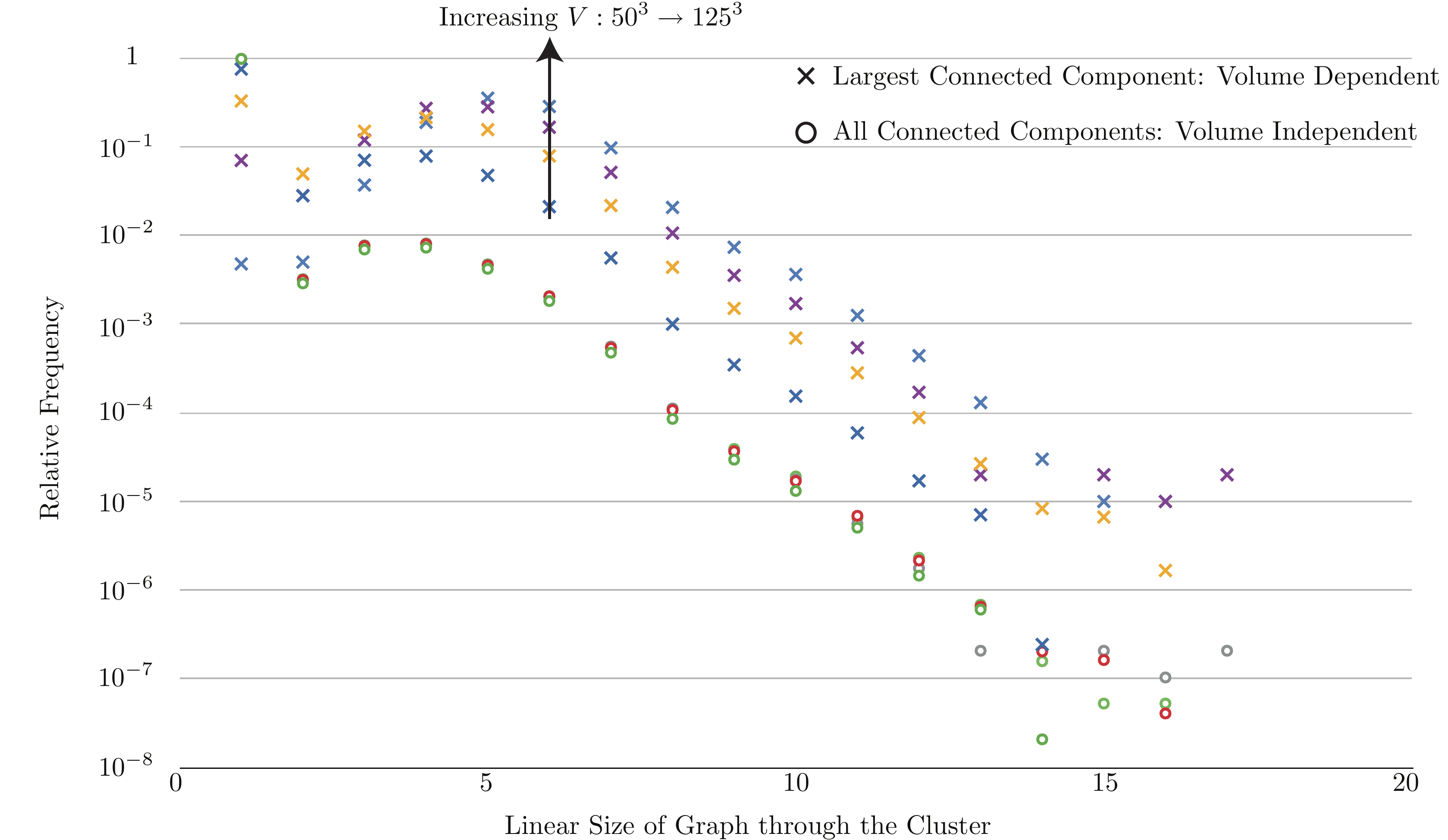}}
\end{center}
\vspace*{-10pt}
\caption{Volume independence of distribution of {\em all} connected components within a 
cluster volume.  The upper four curves are simulations calculating the {\em largest} connected 
component in cluster volumes of $V = (50^3,75^3,100^3,125^3)$ while the 
lower four curves examine the distribution for {\em all} connected 
components at $p=10^{-4}$ and $m_e=6$ (total number of 
simulations vary between $O(10^5)-O(10^6)$).  As you can see, volume independence when 
calculating all connected components is good.  Additionally, we are now calculating the 
maximum {\em physical} separation of endpoints within each connected component.  This 
has the effect of smoothing out the curves in Fig.~\ref{figure:peaks2} and slightly 
shifting the main peak to the left 
(as maximum graph diameter, in general, is larger than the maximum 
node separation in the physical cluster). In the simulations calculating the largest 
connected component, the main peaks shift to the right as volume increases.  This is 
again due to the fact that the {\em largest} connected component will scale with volume. }
\label{figure:peaks3}
\end{figure*}

\bibliographystyle{apsrev}
\bibliography{bib1}  

%
%

%

\end{document}